\documentclass[11pt]{article}
\usepackage{amsmath}
\usepackage{amssymb}
\usepackage{theorem}
\usepackage{booktabs}
\usepackage{epsfig}
\usepackage{graphicx}
\usepackage{epstopdf}
\usepackage{centernot}
\usepackage[algoruled,linesnumbered,noend]{algorithm2e}

\DeclareMathOperator{\Pref}{Pref}
\DeclareMathOperator{\Card}{Card}
\DeclareMathOperator{\merge}{merge}
\DeclareMathOperator{\LAST}{LAST}

\newtheorem{proposition}{Proposition}[section]

\newtheorem{definition}{Definition}[section]
\newtheorem{theorem}{Theorem}[section]

\newtheorem{lemma}{Lemma}[section]
\newtheorem{remark}{Remark}[section]
\newtheorem{example}{Example}[section]
\newtheorem{corollary}{Corollary}[section]

\newcommand{\N}{\mathbb{N}}

\newcommand{\Li}{\mathcal{L}}
\newcommand{\Fi}{\mathcal{F}}
\newcommand{\PMCI}{\mathcal{PMC}}
\newcommand{\PAL}{\mathcal{P}_{aL}}
\newcommand{\SP}{\mathcal{SP}}

\def\petitcarre{\vrule height4pt width 4pt depth0pt}
\def\enddim{\relax\ifmmode\eqno{\hbox{\petitcarre}}
\else
{\unskip\nobreak\hfil\penalty50
   \hskip2em\hbox{}\nobreak\hfil
   \petitcarre
   \parfillskip=0pt \finalhyphendemerits=0
  \par\medskip}\fi}

\def \begdim {\noindent {\sc Proof} : \par \noindent}

\DeclareMathOperator{\CFL}{CFL}
\DeclareMathOperator{\ICFL}{ICFL}

\DeclareMathOperator{\al}{alph}

\numberwithin{equation}{section}

\title{\Large \bf Inverse Lyndon words and
Inverse Lyndon factorizations of words}
\author{Paola Bonizzoni$^1$, Clelia De Felice$^2$, Rocco Zaccagnino$^2$, Rosalba Zizza$^2$ \\
$^1$Universit\`a degli Studi di Milano Bicocca, $^2$Universit\`a degli Studi di Salerno}

\begin{document}

\maketitle

\thispagestyle{plain}

\begin{abstract}
Motivated by applications to string processing, we introduce variants of the Lyndon
factorization called inverse Lyndon factorizations. Their factors, named inverse Lyndon words, are in a class
that strictly contains anti-Lyndon words, that is Lyndon words with respect to the inverse lexicographic order.
The Lyndon factorization of a nonempty word $w$ is unique but $w$ may have several inverse Lyndon factorizations.
We prove that any nonempty word $w$ admits a canonical inverse Lyndon factorization, named $\ICFL(w)$, that maintains
the main properties of the Lyndon factorization of $w$: it
can be computed in linear time, it is uniquely determined, it preserves a compatibility property
for sorting suffixes. In particular, the compatibility property of $\ICFL(w)$ is a consequence of another
result: any factor in $\ICFL(w)$ is a concatenation of consecutive factors of the
Lyndon factorization of $w$ with respect to the inverse lexicographic order.
\end{abstract}

\section{Introduction} \label{intro}

Lyndon words were introduced in \cite{Lyndon0}, as {\it standard lexicographic sequences},
and then used in the context of the free groups in \cite{Lyndon}.
A Lyndon word is a word which is strictly smaller than each of its proper
cyclic shifts for the lexicographical ordering.
A famous theorem concerning Lyndon words asserts that any nonempty word
factorizes uniquely as a nonincreasing product of Lyndon words, called its
Lyndon factorization. This theorem, that can be recovered from results in \cite{Lyndon},
provides an example of a factorization of a free monoid, as defined in \cite{sch}
(see also \cite{bpr,Lo}). Moreover, there are several results which give relations between
Lyndon words, codes and combinatorics of words \cite{bepe}.

The Lyndon factorization has recently revealed to be a useful tool also in string processing
algorithms \cite{Bannai15,MM13} with
strong potentialities that have not been completely explored and understood.
This is due also to the fact that it can be efficiently computed.
Linear-time algorithms for computing this factorization
can be found in \cite{duval,FM} whereas an $\mathcal{O}(\lg{n})$-time parallel algorithm
has been proposed in \cite{apostolico-crochemore-1995,daykin}.
A connection between the Lyndon factorization and
the Lempel-Ziv (LZ) factorization has been given in \cite{Karkai17}, where it is shown that in general
the size of the LZ factorization is larger than the size of the
Lyndon factorization, and in any case the size of the Lyndon factorization cannot be
larger than a factor of 2 with respect to the size of LZ.

Relations between Lyndon words and the Burrows-Wheeler Transform (BWT)
have been discovered first in \cite{CDP,MRRS} and, more recently, in \cite{varBWT}.
Variants of BWT proposed in the previous papers are based on combinatorial results
proved in \cite{GeReu} (see \cite{PerRes} for further details and \cite{GeResReu}
for more recent related results).

Lyndon words are lexicographically smaller than all its proper nonempty suffixes.
This explains why the Lyndon factorization has become of particular
interest also in suffix sorting problems.
The suffix array (SA) of a word $w$ is the lexicographically ordered
list of the starting positions of the suffixes of $w$.
The connection between Lyndon factorizations and suffix arrays
has been pointed out in \cite{from-SA-to-Lyndon},
where the authors show a method to construct the Lyndon factorization of a text from its SA.
Conversely, the computation of the SA of a text from its Lyndon factorization has been
proposed in \cite{from-lyndon-to-SA}
and then explored in \cite{restivo-sorting,restivo-sorting-2014}.

The algorithm proposed in \cite{restivo-sorting,restivo-sorting-2014}
is based on the following interesting combinatorial result,
proved in the same papers:
if $u$ is a concatenation of consecutive Lyndon factors
of $w = xuy$, then the position of a nonempty suffix $u_i$ in
the ordered list of suffixes of $u$ (called {\em local} suffixes)
is the same position of the nonempty suffix $u_iy$ in the ordered list
of the suffixes of $w$ (called {\em global} suffixes).
In turn, this result suggests a divide and conquer strategy for the sorting of the suffixes of a word $w = w_1w_2$:
we order the nonempty suffixes of $w_1$ and the nonempty suffixes of $w_2$ independently (or in parallel) and then
we merge the resulting lists (see Section \ref{restivo-sorting} for further details).

However, in order to have a practical interest, the divide and conquer approach
proposed in \cite{restivo-sorting-2014}
would require two main ingredients:
an efficient algorithm to perform the merging of two sorted lists, which is still to be improved,
and a strategy to manage the size of the factors in a Lyndon factorization.
Indeed, we may have extreme cases of very short or very long factors (a word $a^k$ is factorized into factors of size $1$, while we may
have Lyndon words of huge size).
On the other hand LZ may produce factorizations of very large sizes which are not useful for a good compression
(as in the case of genomic sequences).

In this paper we face the following main question, raised by the above discussion:
{\em can we define a factorization of $w$,
which maintains some useful properties of the Lyndon factorization
but that allows us to manage the size of the factors?}
We first introduce the notion of
an {\it inverse Lyndon word}, that is a word greater than any of its proper nonempty suffixes
(Section \ref{ILFsection}).
We compare this notion with that of Lyndon words with respect to the inverse lexicographic order
(or anti-Lyndon words \cite{antiLyndon}) and with
the notion of strict sesquipowers of Lyndon
words (see \cite{duval}). We show that
the set of the inverse Lyndon words is equal to the set
of the strict sesquipowers of anti-Lyndon words and, consequently, it
strictly contains the class of the
anti-Lyndon words.
Then we give the definition of an inverse Lyndon factorizations of a word,
whose factors are inverse Lyndon words (Section \ref{ILFsection}).
The Lyndon factorization of a nonempty word $w$ is unique but $w$ may
have several inverse Lyndon factorizations.
As a main result, we define a
canonical inverse Lyndon factorization of a nonempty word $w$, denoted
by $\ICFL(w)$. We prove that $\ICFL(w)$ can be still computed in linear time
and it is uniquely determined.
Moreover, if $w$ is a Lyndon word different from a letter,
then $\ICFL(w)$ has at least two factors and a converse holds for an inverse Lyndon
word and its Lyndon factorization.
Finally, we prove that $\ICFL(w)$ belongs to a special class of inverse Lyndon factorizations,
called {\it groupings} (see Section \ref{groupings}), and then each of its factors is a
concatenation of consecutive factors of
the Lyndon factorization of $w$ with respect to the inverse lexicographic order.
Hence the compatibility property proved in \cite{restivo-sorting}
applies also to $\ICFL(w)$, with respect to the inverse lexicographic order.

In order to answer the above question, we propose to combine the Lyndon factorization of a word $w$ with $\ICFL(w)$.
We test our proposal by running an experimental analysis over two biological datasets
\footnote{http://www.di.unisa.it/professori/zizza/EXP/experiments.pdf}.
Experiments confirm that we obtain a factorization of intermediate size
between that of LZ and that of the Lyndon factorization.

The paper is organized as follows.
In Section \ref{prel}, we gathered
the basic definitions
and known results we need.
Inverse Lyndon words are discussed in Section \ref{ILW}.
Inverse Lyndon factorizations and $\ICFL(w)$ are presented
in Section \ref{ILFsection}. More precisely, for the construction of $\ICFL(w)$
we need a special prefix of $w$, defined in Section \ref{bre},
whereas we give the recursive definition of the factorization in
Section \ref{icfl}.
A linear-time algorithm for computing $\ICFL(w)$ is presented
in Section \ref{AlgoICFL}. This algorithm uses two subroutines
described in Section \ref{AlgoFindbre}. We introduce groupings in
Section \ref{groupings} and we prove that $\ICFL(w)$ falls in this
class of factorizations in Section \ref{icfl-grouping}.

\section{Preliminaries} \label{prel}

For the material in this section see
\cite{bpr,CK,Lo,lothaire,reu}.

\subsection{Words}

Let $\Sigma^{*}$ be the {\it free monoid}
generated by a finite alphabet $\Sigma$
and let $\Sigma^+=\Sigma^{*} \setminus 1$, where $1$ is
the empty word.
For a set $X$, $\Card(X)$ denotes the cardinality of $X$.
For a word $w \in \Sigma^*$, we denote by $|w|$ its {\it length}.
A word $x \in \Sigma^*$ is a {\it factor} of $w \in \Sigma^*$ if there are
$u_1,u_2 \in \Sigma^*$ such that $w=u_1xu_2$.
If $u_1 = 1$ (resp. $u_2 = 1$), then $x$ is a {\it prefix}
(resp. {\it suffix}) of $w$.
A factor (resp. prefix, suffix) $x$ of $w$
is {\it proper} if $x \not = w$.
Two words $x,y$ are {\it incomparable} for the prefix order, and we write $x \Join y$,
if neither $x$ is a prefix of $y$ nor $y$ is a prefix of $x$.
Otherwise, $x,y$ are {\it comparable} for the prefix order.
We write $x \leq_{p} y$ if $x$ is a prefix of $y$
and  $x \geq_{p} y$ if $y$ is a prefix of $x$.
The following result, named {\it overlapping-suffix lemma} in \cite{CLR},
is a direct consequence of the definitions.

\begin{lemma} \label{basic4}
Let $x,y,w \in \Sigma^+$ such that $x$ and $y$ are both prefixes
(resp. suffixes) of $w$. If $|x| <  |y|$,
then $x$ is a proper prefix (resp. proper suffix) of $y$.
If $|x| = |y|$, then $x = y$.
Conversely, let $x$ be a prefix (resp. a suffix) of $y$
and let $y$ be a prefix (resp. a suffix) of $w$. If $|x| <  |y|$, then
$x$ is a proper prefix (resp. a proper suffix) of $w$.
\end{lemma}

We recall that two words $x,y$ are called {\em conjugate} if there exist words
$u,v$ such that $x=uv, y=vu$.
The conjugacy relation is an equivalence relation. A conjugacy class
(or necklace) is a class of this equivalence relation.
The following is Proposition 1.3.4 in \cite{lotVecchio}.

\begin{proposition} \label{equazione}
Two words $x,y \in \Sigma^+$ are conjugate if and only if there exists
$z \in \Sigma^*$ such that
\begin{eqnarray} \label{EQ1}
xz & = & zy
\end{eqnarray}
More precisely, equality (\ref{EQ1}) holds if and only if there exist
$u,v \in \Sigma^*$ such that
\begin{eqnarray} \label{EQ2}
x & = & uv, \quad y = vu, \quad z \in u(vu)^*.
\end{eqnarray}
\end{proposition}

A {\it sesquipower} of a word $x$ is a word $w = x^np$ where
$p$ is a proper prefix of $x$ and $n \geq 1$.
A nonempty word $w$ is
\textit{unbordered} if
no proper nonempty prefix of $w$ is a suffix of $w$.
Otherwise, $w$ is {\it bordered}.
A nonempty word $w$ is \textit{primitive} if
$w = x^k$ implies $k = 1$. An unbordered word is primitive.
Let $r,w$ be nonempty words over $\Sigma$. We say that two occurrences of $r$ as a factor of $w$
{\it overlap} if $w = xrz = x'rz'$ with $|x'| < |x| < |x'r|$. Therefore $r$ is
bordered. The following lemma will be used in Section \ref{AlgoFindbre}.

\begin{lemma} \label{basic3}
Let $x,y,w,r \in \Sigma^+$ be such that
$$w = xr = ry,$$
with $|x| < |r|$,
i.e., $r$ occurs twice in $w$ and these two occurrences of $r$ in $w$ overlap.
Then there exists $r' \in \Sigma^+$ such that
$$w = x'r' = r'y',$$
with $|r'| < |x'|$, and $y',y$ start with the same letter.
\end{lemma}
\begdim
Let $x,y,r \in \Sigma^+$ be as in the statement.
By Proposition \ref{equazione}, there are $u,v \in \Sigma^*$ and $n \in \N$ such that
\begin{eqnarray} \label{EQ6}
x & = & uv, \quad y = vu, \quad r = u(vu)^n.
\end{eqnarray}
Set
\begin{eqnarray} \label{EQ7}
r' = \begin{cases} u & \mbox{ if } u \not = 1 \\
v & \mbox{ if }  u  = 1 \end{cases}
\end{eqnarray}
and
\begin{eqnarray} \label{EQ8}
x' = \begin{cases} x(uv)^n = (uv)^{n+1} & \mbox{ if } u \not = 1 \\
v^n & \mbox{ if }  u  = 1  \end{cases} \quad
y' = \begin{cases} (vu)^ny = (vu)^{n+1} & \mbox{ if } u \not = 1 \\
v^n & \mbox{ if }  u  = 1  \end{cases}
\end{eqnarray}
By using Eqs. (\ref{EQ6})-(\ref{EQ8}), we can easily see that
$x'r' = r'y' = ry = xr$.
Moreover, since $|x| < |r|$, if $u \not = 1$, then $n \geq 1$ and
if $u = 1$, then $n \geq 2$. Hence, in both cases, $|r'| < |x'|$.
Finally, $y$ is a prefix of $y'$, therefore they start with the same letter.
\enddim

\subsection{Lexicographic order and Lyndon words}

\begin{definition} \label{lex-order}
Let $(\Sigma, <)$ be a totally ordered alphabet.
The {\it lexicographic} (or {\it alphabetic order})
$\prec$ on $(\Sigma^*, <)$ is defined by setting $x \prec y$ if
\begin{itemize}
\item $x$ is a proper prefix of $y$,
or
\item $x = ras$, $y =rbt$, $a < b$, for $a,b \in \Sigma$ and $r,s,t \in \Sigma^*$.
\end{itemize}
\end{definition}

For two nonempty words $x,y$, we write $x \ll y$ if
$x \prec y$ and $x$ is not a proper prefix of $y$
\cite{Bannai15}. We also write $y \succ x$ if $x \prec y$.
Basic properties of the lexicographic order are recalled below.

\begin{lemma} \label{proplexord}
For $x,y \in \Sigma^*$, the following properties hold.
\begin{itemize}
\item[(1)]
$x \prec y$ if and only if $zx \prec zy$,
for every word $z$.
\item[(2)]
If $x \ll y$, then $xu \ll yv$
for all words $u,v$.
\item[(3)]
If $x \prec y \prec xz$ for a word $z$,
then $y = xy'$ for some word $y'$ such that $y' \prec z$.
\end{itemize}
\end{lemma}

\begin{definition}\label{Lyndon-word}
A Lyndon word $w \in \Sigma^+$ is a word which is primitive and the smallest one
in its conjugacy class for the lexicographic order.
\end{definition}

\begin{example}
{\rm Let $\Sigma = \{a,b\}$ with $a < b$.
The words $a$, $b$, $aaab$, $abbb$, $aabab$ and $aababaabb$
are Lyndon words. On the contrary, $aba$ and
$abaab$ are not Lyndon words. Indeed, $aab \prec aba$
and $aabab \prec abaab$.}
\end{example}

Lyndon words are also called {\it prime words}
and their prefixes are also called {\it preprime words} in \cite{Knuth}.
Interesting properties of Lyndon words are recalled below.

\begin{proposition} \label{P1}
Each Lyndon word $w$ is unbordered.
\end{proposition}

\begin{proposition} \label{P2}
A word $w \in \Sigma^+$ is a Lyndon word if and only if
$w \prec s$, for each nonempty proper suffix $s$ of $w$.
\end{proposition}

\subsection{The Lyndon factorization} \label{LyFa}

A family $(X_i)_{i \in I}$ of subsets of $\Sigma^+$, indexed by a totally ordered set $I$, is
a {\it factorization of the free monoid} $\Sigma^*$ if each word $w \in \Sigma^*$
has a unique factorization
$w = x_1 \cdots x_n$, with $n \geq 0$, $x_i \in X_{j_i}$ and $j_1 \geq j_2 \ldots \geq j_n$ \cite{bpr}.
A factorization $(X_i)_{i \in I}$ is called {\it complete} if each $X_i$ is reduced to a singleton
$x_i$ \cite{bpr}.
Let $L = L_{(\Sigma^*, <)}$ be the set of Lyndon words, totally
ordered by the relation $\prec$ on $(\Sigma^*, <)$.
The following theorem,
shows that the family $(\ell)_{\ell \in L}$ is a complete factorization of $\Sigma^*$.

\begin{theorem} \label{Lyndon-factorization}
Any word $w \in \Sigma^+$ can be written in a unique way as
a nonincreasing product $w=\ell_1 \ell_2 \cdots \ell_h$ of Lyndon words, i.e., in the form
\begin{eqnarray} \label{LF}
w & = & \ell_1 \ell_2 \cdots \ell_h, \mbox{ with } \ell_j \in L \mbox{ and } \ell_1 \succeq \ell_2 \succeq \ldots \succeq \ell_h
\end{eqnarray}
\end{theorem}

The sequence $\CFL(w) = (\ell_1, \ldots, \ell_h)$ in Eq. (\ref{LF}) is called the
\textit{Lyndon decomposition} (or \textit{Lyndon factorization}) of $w$.
It is denoted by $\CFL(w)$ because Theorem \ref{Lyndon-factorization}
is usually credited to Chen, Fox and Lyndon
\cite{Lyndon}.
Uniqueness of the above factorization is a consequence of the following result, proved
in \cite{duval}.

\begin{lemma} \label{duval-prop}
Let $w \in \Sigma^+$ and let $\CFL(w) = (\ell_1, \ldots, \ell_h)$.
Then the following properties hold:
\begin{itemize}
\item[(i)]
$\ell_h$ is the nonempty suffix of $w$ which is the smallest with respect to the lexicographic order.
\item[(ii)]
$\ell_h$ is the longest
suffix of $w$ which is a Lyndon word.
\item[(iii)]
$\ell_1$ is the longest
prefix of $w$ which is a Lyndon word.
\end{itemize}
\end{lemma}

Therefore, given $w \in \Sigma^+$, if $\ell_1$ is its longest
prefix which is a Lyndon word and $w= \ell_1 w'$, then $\CFL(w) = (\ell_1, \CFL(w'))$.
As a consequence of Theorem \ref{Lyndon-factorization}, for any word $w$ there is a factorization
$$w = \ell_1^{n_1} \cdots \ell_r^{n_r}$$
where $r  > 0$, $n_1, \ldots , n_r \geq 1$, and $\ell_1 \succ \ldots \succ \ell_r$
are Lyndon words, also named {\it Lyndon factors} of $w$.
There is a linear time algorithm
to compute the pair $(\ell_1, n_1)$ and thus, by iteration,
the Lyndon factorization of $w$.
It is due to Fredricksen and Maiorana \cite{FM} and it is also
reported in \cite{lothaire}.
It can also be used to compute the Lyndon word in the conjugacy class of a primitive
word in linear time \cite{lothaire}.
Linear time algorithms may also be found in \cite{duval} and in the more recent paper
\cite{alternative}.

\subsection{Inverse lexicographic order and anti-Lyndon words} \label{invlo}

We also need the following well-known definition.

\begin{definition} \label{ILO}
Let $(\Sigma, <)$ be a totally ordered alphabet. Let $<_{in}$ be the inverse of $<$, defined
by
$$ \forall a, b \in \Sigma \quad b <_{in} a \Leftrightarrow a < b $$
The {\rm inverse lexicographic} or {\rm inverse alphabetic order},
denoted $\prec_{in}$, on $(\Sigma^*, <)$ is the lexicographic order
on $(\Sigma^*, <_{in})$.
\end{definition}

From now on, $L_{in} = L_{(\Sigma^*, <_{in})}$ denotes the set
of the Lyndon words on $\Sigma^*$ with respect to the inverse lexicographic order.
A word $w \in L_{in}$ will be named an {\it anti-Lyndon word}.

\subsection{Sorting suffixes of a text} \label{restivo-sorting}

In \cite{restivo-sorting,restivo-sorting-2014}, the authors
found interesting relations between the sorting of the suffixes of a word $w$
and that of its factors.
Let $w, x, u, y \in \Sigma^*$, and let $u$ be a nonempty factor of $w = xuy$.
Let $first(u)$ and $last(u)$ denote
the position of the first and the last symbol of $u$ in $w$, respectively.
If $w=a_1 \cdots a_n$, $a_i \in \Sigma$, $1 \leq i \leq j \leq n$,
then we also set $w[i, j]=a_i \cdots a_j$.
A \textit{local suffix} of $w$ is a suffix of a factor of $w$, specifically
$suf_u(i) = w[i, last(u)]$ denotes the \textit{local suffix} of $w$
at the position $i$ with respect to $u$, $i \geq first(u)$.
The corresponding \textit{global suffix} $suf_u(i)y$ of
$w$ at the position $i$ is denoted by
$suf_w(i) = w[i, last(w)]$ (or simply $suf(i)$ when it is understood).
We say that $suf_u(i)y$ is {\it associated} with $suf_u(i)$.

\begin{definition} \cite{restivo-sorting,restivo-sorting-2014} \label{def-sorting}
Let $w \in \Sigma^+$ and let $u$ be a factor of $w$.
We say that the sorting of the nonempty local suffixes of
$w$ with respect to $u$ is {\rm compatible}
with the sorting of the corresponding nonempty global suffixes of $w$
if for all $i,j$ with $first(u) \leq i < j \leq last(u)$,
$$suf_u(i) \prec suf_u(j) \Longleftrightarrow suf(i) \prec suf(j).$$
\end{definition}

Let $u = \ell_r \cdots \ell_s$ be a concatenation of consecutive Lyndon factors of
$w$.
Let $\Li_{\mathit loc}(u,w) = (s_1, \ldots, s_t)$ be the ordered list of the
nonempty local suffixes of $w$ with respect to $u$ and let
$\Li_{\mathit glob}(u,w) =(s'_1, \ldots, s'_t)$ be the ordered
list of the corresponding nonempty global suffixes of $w$.
We name it the {\it global list associated} with $\Li_{\mathit loc}(u,w)$.
The following result proved in \cite{restivo-sorting,restivo-sorting-2014}
shows that each $s'_i$ in
$\Li_{\mathit glob}(u,w)$ is associated with $s_i$.

\begin{theorem} \label{teo-sorting}
Let $w \in \Sigma^+$ and let $\CFL(w) = (\ell_1, \ldots, \ell_h)$ be
its Lyndon factorization.
Then, for any $r,s$, $1 \leq r \leq s \leq h$,
the sorting of the nonempty local suffixes of
$w$ with respect to $u = \ell_r \cdots \ell_s$
is compatible with the sorting of the corresponding nonempty global suffixes of $w$.
\end{theorem}

\begin{remark}
{\rm The compatibility property stated in Theorem
\ref{teo-sorting} does not hold for the
empty suffix of a word. Indeed, let $\Sigma =\{ a, b, c, d \}$ with $a < b < c < d$.
Let $w = bbcbcacad$, thus $\CFL(w) = (bbc, bc, acad)$.
Consider the local suffixes $1, bbc$ of $w$ with respect to $u = bbc$
and the corresponding global suffixes $bcacad, w$. We have $1 \prec bbc$ but
$w \prec bcacad$.}
\end{remark}

If $\Li_1$ and $\Li_2$ are
two sorted lists of elements taken from any totally ordered set, then
the result of the operation $\merge(\Li_1,\Li_2)$ is a single sorted list containing the
elements of $\Li_1$ and $\Li_2$.
Theorem \ref{teo-sorting} could be considered in a merge sort algorithm
for the sorting of the suffixes of $w$,
as suggested in \cite{restivo-sorting,restivo-sorting-2014}.
Starting with the list $\CFL(w) = (\ell_1, \ldots, \ell_h)$,
it could operate as follows.
\begin{itemize}
\item[-]
Divide the sequence into two subsequences
$(\ell_1, \ldots, \ell_r)$, $(\ell_{r+1}, \ldots, \ell_h)$,
where $r = \lceil k/2 \rceil$
\item[-]
Let $\Li_{1}$ be the list of the nonempty suffixes of $u = \ell_1 \cdots \ell_r$,
let $\Li_{2}$ be the list of the nonempty suffixes of $y = \ell_{r+1} \cdots \ell_h$.
Sort the two subsequences $\Li_{1}, \Li_{2}$
recursively using merge sort, thus
obtaining $\Li_{\mathit loc}(u,w), \Li_{\mathit loc}(y,w)$
\item[-]
Merge the two subsequences
$\Li_{\mathit glob}(u,w), \Li_{\mathit glob}(y,w)$
to produce $\Li_{\mathit glob}(w,w)$.
\end{itemize}
Notice that in the third step we change $\Li_{\mathit loc}(u,w), \Li_{\mathit loc}(y,w)$
into $\Li_{\mathit glob}(u,w), \Li_{\mathit glob}(y,w)$. Indeed,
as pointed out in Example \ref{Merge},
two problems arise if we considered the local lists in the merge step.
First, there could exist two local suffixes $s,s'$, with
corresponding global suffixes $sz, s'z'$,
such that $s \prec s'$ but $s'z' \prec sz$.
Furthermore, an element $s$ could occur
twice in $\merge(\Li_{\mathit loc}(u,w), \Li_{\mathit loc}(y,w))$.
In this case, when we produce the global list
$\Li_{\mathit glob}(w,w)$,
we have to change the second occurrence
of $s$ into $sy$.

\begin{example} \label{Merge}
{\rm Let $\Sigma =\{ a, b, c, d \}$ with $a < b < c < d$.
Let $w = bbcbcacad$, thus $\CFL(w) = (bbc, bc, acad)$. Then,
$\Li_{\mathit loc}(bbc, w) = (bbc, bc, c)$,
$\Li_{\mathit loc}(bc, w) = (bc, c)$,
and $\Li_{\mathit loc}(acad,w) = (acad, ad, cad, d)$.
Consider the local suffix $c$ in $\Li_{\mathit loc}(bbc, w)$,
the local suffix $cad$ in $\Li_{\mathit loc}(acad,w)$ and the corresponding associated
global suffixes $cbcacad$ and $cad$.
We see that $c \prec cad$ but $cbcacad \succ cad$.
Let $u = bbcbc$, $y = acad$, then
\begin{eqnarray*}
\Li_{\mathit loc}(bbcbc, w) & = & \Li_{\mathit glob}(bbcbc, u) \\
& = & \merge (\Li_{\mathit glob}(bbc, u), \Li_{\mathit glob}(bc, u)) \\
& = & \merge ((bbcbc, bcbc, cbc), (bc, c)) = (bbcbc, bc, bcbc, c, cbc) \\
\Li_{\mathit glob}(w,w) & = & \merge (\Li_{\mathit glob}(u, w), \Li_{\mathit glob}(y, w)) \\
& = &\merge ((bbcbcacad, bcacad, bcbcacad, cacad, cbcacad), (acad, ad, cad, d)) \\
& = & (acad, ad, bbcbcacad, bcacad, bcbcacad, cacad, cad, cbcacad, d)
\end{eqnarray*}
If in the third step we merged $\Li_{\mathit loc}(bbc, u),
\Li_{\mathit loc}(bc, u)$, we would obtain $(bbc, bc, bc, c, c)$.
Then, in order to
obtain $\Li_{\mathit loc}(bbcbc, w)$, we have to change $bbc$ into $bbcbc$, the second occurrence
of $bc$ into $bcbc$ and the second occurrence of $c$ into $cbc$.}
\end{example}

\section{Inverse Lyndon words} \label{ILW}

As mentioned in Section \ref{intro}, the
Lyndon factorization of a word may
generate very long or too short factors, thus becoming unsatisfactory with respect to
a parallel strategy.
We face this problem in Section \ref{ILFsection}, where we
introduce another factorization which maintains the main properties of the
Lyndon factorization but that allows us to overcome the glitch.
This factorization is based on the notion of inverse Lyndon words, given below.

\begin{definition} \label{inverse-Lyndon-word}
A word $w \in \Sigma^+$ is an inverse Lyndon word if
$s \prec w$, for each nonempty proper suffix $s$ of $w$.
\end{definition}

\begin{example}
{\rm The words $a$, $b$, $bbba$, $baaab$, $bbaba$ and $bbababbaa$
are inverse Lyndon words on $\{a,b\}$, with $a < b$.
On the contrary, $aaba$ is not an inverse Lyndon word since $aaba \prec ba$.
Analogously, $aabba \prec ba$ and thus $aabba$ is not an inverse Lyndon word.}
\end{example}

In Section \ref{inv} we will see that
the set of inverse Lyndon words properly contains the set of Lyndon words with respect to
the inverse lexicographic order (or anti-Lyndon words). Then, in Section \ref{SSA}
we will prove
that the set of the inverse Lyndon words is equal to the set
of the strict sesquipowers of anti-Lyndon words.
Some useful properties of the inverse Lyndon words are proved below.
The following is a direct consequence of Definitions \ref{lex-order}, \ref{inverse-Lyndon-word}.

\begin{lemma} \label{notinv}
If $w \in \Sigma^+$ is not an inverse Lyndon word, then there exists a
nonempty proper suffix $s$ of $w$ such that $w \ll s$.
\end{lemma}

Next lemma shows that the set of the inverse Lyndon words (with the empty word)
is a {\it prefix-closed} set, that is, it contains the prefixes of its elements.

\begin{lemma} \label{lem:inverse-Lyndon-word-prefix}
Any nonempty prefix of an
inverse Lyndon word is an inverse Lyndon word.
\end{lemma}
\begdim
Let $w \in \Sigma^+$ be an inverse Lyndon word.
By contradiction, assume that there is a proper nonempty prefix
$p$ of $w = ps$ which is not an inverse Lyndon word.
By Lemma \ref{notinv}, there is
a proper nonempty suffix $v$ of $p$ such that
$p \ll v$. Hence, by item (2) in Lemma \ref{proplexord},
$w = ps \ll vs$.
Thus $w$ is smaller than its proper nonempty suffix $vs$, in
contradiction with the hypothesis that $w$ is
an inverse Lyndon word.
\enddim

Lemmas \ref{lemintermedio} and \ref{lem:inverse-pref-2} are needed for the definition of
our new factorization of a word.

\begin{lemma} \label{lemintermedio}
If $w \in \Sigma^+$ is not an inverse Lyndon word, then there exists a
nonempty proper prefix $p$ of $w = ps$ such that $p \ll s$.
\end{lemma}
\begdim
Let $w$ be as in the statement. By Lemma \ref{notinv},
there exists a shortest nonempty proper suffix $s$ of $w = ps$ such that $w \ll s$.
Hence, $p$ is nonempty. Moreover
there are words $r, t, q \in \Sigma^*$ and letters $a,b \in \Sigma$, with $a < b$,
such that
\begin{eqnarray} \label{EQ3}
s &=& rbq, \quad w = rat = prbq
\end{eqnarray}
We show that $p \ll s$.
If $|p| \geq |ra|$, then $p \ll s$ and the proof is ended.
Thus assume $0 < |p| \leq |r|$.
By Eq.~(\ref{EQ3}), there exists a prefix $t'$ of $t$ such that
$pr = rat'$. Therefore, by Proposition \ref{equazione}, there are $u,v \in \Sigma^*$ such that
\begin{eqnarray} \label{EQ4}
p & = & uv, \quad at' = vu, \quad r \in u(vu)^*.
\end{eqnarray}
If $u = 1$, then $p$ starts with $a$ and so does $w$. Thus $w \ll bq$
and, by the hypothesis on $s$, we have $s = bq$. Hence $r = 1$,
a contradiction.
The same argument applies if $v = 1$.
Thus assume $u \not = 1$, $v \not = 1$. Set $r = u(vu)^k$, $k \geq 0$.
We also have $k > 0$, since, otherwise, by Eq.~(\ref{EQ4}), $p = rv$, with $v \not = 1$,
and consequently $|r| < |p| \leq |r|$, a contradiction.
Finally, by Eqs.~(\ref{EQ3}) and (\ref{EQ4}),
$w = ps = uvs =(uv)^{k+1}ubq$ and $v$ starts with $a$.
Consequently, $w \ll ubq$, with $ubq$ shorter than $s$,
in contradiction with the hypothesis on the length
of $s$.
\enddim

\begin{corollary} \label{ultimo}
A word $w \in \Sigma^+$ is not an inverse Lyndon word if and only if
there are words $r, u, t \in \Sigma^*$ and letters $a,b \in \Sigma$, with $a < b$
such that $w = raurbt$.
\end{corollary}
\begdim
Let $r, u, t \in \Sigma^*$ and $a,b \in \Sigma$, with $a < b$.
By Definition \ref{lex-order} $w = raurbt \ll rbt$, hence
$w$ is not an inverse Lyndon word.
Conversely, assume that $w \in \Sigma^+$ is not an inverse Lyndon word.
By Lemma \ref{lemintermedio}, there exists a
nonempty proper prefix $p$ of $w = ps$ such that $p \ll s$.
By Definition \ref{lex-order}, there are $r, u, t \in \Sigma^*$,
$a,b \in \Sigma$, with $a < b$, such that
$p = rau$, $s = rbt$, and thus
$w = ps = raurbt$.
\enddim

The following lemma shows that there exists $p$ satisfying Lemma \ref{lemintermedio}
and which is in addition an inverse Lyndon word.

\begin{lemma} \label{lem:inverse-pref-2}
If $w \in \Sigma^+$ is not an inverse Lyndon word, then there exists a
nonempty proper prefix $p$ of $w = ps$ such that $p$ is an inverse Lyndon word and $p \ll s$.
\end{lemma}
\begdim
Let $w \in \Sigma^+$ a word which is not an inverse Lyndon word. The proof is
by induction on $|w|$. The shortest nonempty word which is not an inverse Lyndon word
has the form $w = ab$, with $a,b$ letters such that $a < b$.
In this case the nonempty proper prefix $a$ of $w$ is an inverse Lyndon word and $a \ll b = s$.
Now assume $|w| > 2$.
By Lemma \ref{lemintermedio}, there exists a
nonempty proper prefix $p$ of $w = ps$ such that $p \ll s$.
If $p$ is an inverse Lyndon word, then the proof is ended.
Otherwise, by induction hypothesis there exists a
nonempty proper prefix $p'$ of $p = p's'$ such that $p'$ is an inverse Lyndon word and $p' \ll s'$.
Hence, by item (2) in Lemma \ref{proplexord}, $p' \ll s's$.
Thus, $p'$ is a nonempty proper prefix of $w = p's's$ such that $p'$ is an inverse Lyndon word
and $p' \ll s's$.
\enddim

\section{Variants of the Lyndon factorization} \label{ILFsection}

\subsection{Inverse Lyndon factorizations}

We give below the notion of an \textit{inverse Lyndon factorization}.

\begin{definition} \label{inverseLynFact}
Let $w \in \Sigma^+$. A sequence $(m_1, \ldots , m_k)$ of words over $\Sigma$
is an {\rm inverse Lyndon factorization} of $w$ if it satisfies the following conditions
\begin{itemize}
\item[(1)]
$w = m_1 \cdots m_k$,
\item[(2)]
for any $j \in \{1, \ldots , k \}$, the word $m_j$ is an inverse Lyndon word,
\item[(3)]
$m_1 \ll m_2 \ll \ldots \ll m_k$.
\end{itemize}
\end{definition}

Example \ref{nonunique} shows that a word may have different
inverse Lyndon factorizations even with a different number of factors.

\begin{example} \label{nonunique}
{\rm Let $\Sigma = \{a,b,c,d \}$ with $a < b < c < d$,
let $w = dabadabdabdadac \in \Sigma^+$.
The two sequences $(daba, dabdab, dadac)$,
$(dabadab, dabda, dac)$ are both inverse Lyndon factorizations of $w$.
Indeed,
$$w = (daba) (dabdab) (dadac) = (dabadab) (dabda) (dac).$$
Moreover, $daba, dabdab, dadac, dabadab, dabda, dac$ are all
inverse Lyndon words. Furthermore,
$$daba \ll dabdab \ll dadac, \quad dabadab \ll dabda \ll dac.$$
As another example, consider the following two factorizations
of $dabdadacddbdc$
$$(dab)(dadacd)(db)(dc) = (dabda)(dac)(ddbdc)$$
It is easy to see that the two sequences
$(dab,dadacd,db,dc)$, $(dabda,dac,ddbdc)$
are both inverse Lyndon factorizations of
$dabdadacddbdc$. The first factorization has four factors whereas the
second one has three factors.}
\end{example}

In Section \ref{LyFa} we have given the definition of
a complete factorization $(x_i)_{i \in I}$ of the free monoid $\Sigma^*$.
By a result of Sch\"{u}tzenberger, if $(x_i)_{i \in I}$ is a complete factorization
of $\Sigma^*$, then the set $X = \{x_i ~|~ i \in I \}$ is a set of
representatives of the primitive conjugacy classes
(see \cite[Corollary 8.1.7] {bpr}). In particular, any $x_i$ is
a primitive word. The fact that a word $w$ may have several different
inverse Lyndon factorizations is a consequence
of this result since an inverse Lyndon word is not necessarily primitive
(take $baba$, with $a < b$, for instance).

However, we focus on a special canonical inverse Lyndon factorization, denoted
by $\ICFL(w)$, which maintains three important features of $\CFL(w)$:
it is uniquely determined (Proposition \ref{unique}),
it can be computed in linear time (Section \ref{AlgoICFL}) and it maintains
the compatibility property of the nonempty suffixes with respect to the inverse lexicographic
order (Theorem \ref{teo-sorting-inverseBIS}).
We give the definition of $\ICFL(w)$ in Section \ref{icfl}.
It is based on the definition of the \textit{bounded right extension} of a prefix of a word,
defined in Section \ref{bre}.

\subsection{The bounded right extension} \label{bre}

The bounded right extension, abbreviated {\it bre}, of a prefix of a word $w$, defined below,
allows us to define the first factor in the inverse Lyndon factorization
$\ICFL(w)$.

\begin{definition} \label{brepref}
Let $w \in \Sigma^+$, let $p$ be an inverse Lyndon word
which is a nonempty proper prefix
of $w = pv$.
The {\rm bounded right extension} $\overline{p}_w$ of $p$
(relatively to $w$), denoted by $\overline{p}$ when it is
understood, is a nonempty prefix of $v$ such that:
\begin{itemize}
\item[(1)]
$\overline{p}$ is an inverse Lyndon word,
\item[(2)]
$pz'$ is an inverse Lyndon word, for each proper nonempty prefix $z'$
of $\overline{p}$,
\item[(3)]
$p \overline{p}$ is not an inverse Lyndon word,
\item[(4)]
$p \ll \overline{p}$.
\end{itemize}
Moreover, we set
$$\Pref_{bre}(w) = \{(p,\overline{p}) ~|~ p \mbox{ is an inverse Lyndon word
which is a nonempty proper prefix of } w \}.$$
\end{definition}

Notice that, given a word $w$, a nonempty proper prefix $p$ of $w = pv$ can have at most
one bounded right extension $\overline{p}_w$. Indeed,
two different prefixes $p_1, p_2$ of $v$ are comparable for the prefix order, say
$p_1 \leq_p p_2$. If $p_2 = \overline{p}_w$, then $pp_1$ is an inverse Lyndon word and
$p_1$ cannot be a bounded right extension of $p$. Analogously, if $p_1 = \overline{p}_w$,
then $p_2$ cannot be a bounded right extension of $p$ because $p_2$ does not satisfy
condition (2) in Definition \ref{brepref}.
Furthermore, the bounded right extension $\overline{p}_w$ of $p$ might not exist.
For instance, let $\Sigma = \{a,b,c \}$ with $a < b < c$.
For the prefix $ba$ of $baababc$, $\overline{ba}$ does not
exist since any nonempty prefix $p'$ of $ababc$ starts with $a$, thus $p' \ll ba$.
On the contrary, for the prefix $baa$ of $baababc$, we have $bab = \overline{baa}$.
As another example, for the prefix $bab$ of $babc$ we have $c = \overline{bab}$
but for the prefix $ba$ of the same word $babc$, $\overline{ba}$ does not
exist.
Moreover, it is clear that if $w$ is a letter, then $\Pref_{bre}(w)$ is empty.
More generally, if $w$ is an inverse Lyndon word,
then $\Pref_{bre}(w)$ is empty.
A more precise result will be proved below.
We will see that the set $\Pref_{bre}(w)$ is either empty or it is a singleton.
In other words, given a word $w$, either there is no prefix of $w$ which has a bounded
right extension or this prefix is unique. This result will be proved
through Lemmas \ref{lemma1}, \ref{breintermedio} and Proposition \ref{unica}.
Lemmas \ref{lemma0}, \ref{lemma2} below show interesting properties of
$\Pref_{bre}(w)$.

\begin{lemma} \label{lemma0}
Let $w \in \Sigma^+$ and let $(p,\overline{p}) \in \Pref_{bre}(w)$.
Then, there are $r, s, t \in \Sigma^*$, $a, b \in \Sigma$, with $a < b$,
such that $p = ras$ and $\overline{p} = rb$.
\end{lemma}
\begdim
Let $w \in \Sigma^+$ and let $(p,\overline{p}) \in \Pref_{bre}(w)$.
By Definition \ref{brepref},
$p \ll \overline{p}$, hence $p = ras$,
$\overline{p} = rbt$, with $r, s, t \in \Sigma^*$, $a, b \in \Sigma$,
$a < b$.
Moreover $t = 1$, otherwise for any proper prefix $t'$ of $t$, the word $z' = rbt'$
would be a proper prefix of $\overline{p}$ such that $p \ll z'$, thus $pz' \ll z'$ and
$pz'$ would not be an inverse Lyndon word, in contradiction with Definition \ref{brepref}.
\enddim

Recall that, if $w \in \Sigma^+$ is not an inverse Lyndon word, then there exists a
nonempty proper prefix $p$ of $w = ps$ such that $p \ll s$
(Lemma \ref{lemintermedio}).

\begin{lemma} \label{lemma2}
Let $w$ be a nonempty word which is not an inverse Lyndon word but
all its proper nonempty prefixes are inverse Lyndon words.
If $p$ is the longest nonempty proper prefix of $w = ps$ such that
$p \ll s$ for the corresponding suffix $s$, then $(p,s) \in \Pref_{bre}(w)$.
In other words, if $(p,s) \in \mathcal{G} = \{(p',s') ~|~ p's' = w, ~ p' \ll s' \}$
and $|p| \geq |p'|$, for any $(p',s') \in \mathcal{G}$,
then $(p,s) \in \Pref_{bre}(w)$.
\end{lemma}
\begdim
Let $w,p,s$ be words as in the statement.
If $s$ is an inverse Lyndon word, then conditions (1)-(4) in
Definition \ref{brepref} are satisfied with $s = \overline{p}$ and we have done.
Assume that $s$ is not an inverse Lyndon word and
set $p = rax$, $s = rby$, where $r, x, y \in \Sigma^*$, $a, b \in \Sigma$,
$a < b$. We notice that $y = 1$, otherwise $prb = raxrb$ would be a nonempty proper prefix of
$w = ps = prby$ which is not an inverse Lyndon word, since
$prb =raxrb \prec rb$, in contradiction with the hypotheses.
Thus $r \not = 1$ (otherwise, $s = b$ would be an inverse Lyndon word).
Moreover, by Lemma \ref{lemintermedio}, there exists a
nonempty proper prefix $q$ of $s = rb = qt$ such that $q \ll t$.
Notice that $q$ is a prefix of $r = qq'$ and $w = pqt$.
Hence, by item (2) in Lemma \ref{proplexord} and $q \ll t$,
we get $pq = qq'axq \ll t$, with $pq$ longer than $p$,
a contradiction.
\enddim

\begin{lemma} \label{lemma1}
Let $w \in \Sigma^+$. For any nonempty prefix $x$ of $w$, we have
$\Pref_{bre}(x) \subseteq \Pref_{bre}(w)$.
\end{lemma}
\begdim
Let $w \in \Sigma^+$ and let $x$ be a proper nonempty prefix of $w = xv$.
If $\Pref_{bre}(x) = \emptyset$, then the proof is ended.
Otherwise, let $(p,\overline{p}_x) \in \Pref_{bre}(x)$.
By Definition \ref{brepref}, $p$ is a nonempty prefix
of $x = pv'$, hence of $w = pv'v$, which is an inverse Lyndon word.
Moreover, $\overline{p}_x$ is a nonempty prefix of $v'$, thus of $v'v$, such that
conditions (1)-(4) holds.
Therefore, $(p,\overline{p}_x) = (p,\overline{p}_w) \in \Pref_{bre}(w)$.
\enddim

\begin{lemma} \label{breintermedio}
A word $w \in \Sigma^+$ is not an inverse Lyndon word if and only if
$\Pref_{bre}(w)$ is nonempty.
\end{lemma}
\begdim
Let $w \in \Sigma^+$ be a word such that
$\Pref_{bre}(w) \not = \emptyset$ and let $(p,\overline{p}) \in \Pref_{bre}(w)$.
By Definition \ref{brepref}, the nonempty
prefix $p \overline{p}$ of $w$ is not an inverse Lyndon word,
thus $w$ is not an inverse Lyndon word by Lemma \ref{lem:inverse-Lyndon-word-prefix}.

Conversely, let $w \in \Sigma^+$ be a word which is not an inverse Lyndon word.
We prove that $\Pref_{bre}(w)$ is nonempty by induction on $|w|$.
The shortest nonempty word which is not an inverse Lyndon word
has the form $w = ab$, with $a,b$ letters such that $a < b$
and $\Pref_{bre}(ab) = \{(a,b)\}$.

Assume $|w| > 2$ and let $p'$ be the shortest nonempty prefix of $w$
which is not an inverse Lyndon word. Hence $|p'| \geq 2$.
If $|p'| < |w|$, then $\emptyset \not = \Pref_{bre}(p') \subseteq \Pref_{bre}(w)$
(induction hypothesis and Lemma \ref{lemma1}).
Otherwise, $p' = w$ and all nonempty proper prefixes of $w$ are inverse Lyndon words.
Thus, by Lemma \ref{lemintermedio}, there exists a
nonempty proper prefix $p$ of $w = ps$ such that $p \ll s$. Choose it of maximal length.
By Lemma \ref{lemma2},  $(p,s) \in \Pref_{bre}(w)$.
\enddim

\begin{proposition} \label{unica}
If $w \in \Sigma^+$ is not an inverse Lyndon word,
then $\Card(\Pref_{bre}(w)) = 1$.
\end{proposition}
\begdim
Let $w \in \Sigma^+$ be a word which is not an inverse Lyndon word.
By Lemma \ref{breintermedio}, the set $\Pref_{bre}(w)$ is nonempty.
By contradiction, let $(p,\overline{p}), (q,\overline{q}) \in \Pref_{bre}(w)$,
with $p \not = q$.
Since $p\overline{p}$ and $q \overline{q}$ are both prefixes of $w$, they are
comparable for the prefix order and one of the following three cases holds.
\begin{itemize}
\item[(1)]
$p\overline{p}$ is a proper prefix of $q \overline{q}$,
\item[(2)]
$q \overline{q}$ is a proper prefix of $p\overline{p}$,
\item[(3)]
$p\overline{p} = q \overline{q}$.
\end{itemize}
In case (1), either $p\overline{p}$ is a prefix of $q$ or
$p\overline{p} = q z'$, for a proper prefix $z'$ of $\overline{q}$.
Since $p\overline{p}$ is not an inverse Lyndon word, both cases are impossible
(the first contradicts Lemma \ref{lem:inverse-Lyndon-word-prefix}, the second
Definition \ref{brepref}).
We may exclude case (2) by a similar reasoning, thus assume that
$p\overline{p} = q \overline{q}$, with $q$ being a nonempty proper prefix of $p$
(the same argument applies if $p$ is a proper prefix of $q$).
By Lemma \ref{lemma0}, there are $r, s, r', s' \in \Sigma^*$,
$a, b, a', b' \in \Sigma$, with $a < b$, $a' < b'$, such
that $p = ras$, $\overline{p} = rb$, $q = r'a's'$, $\overline{q} = r'b'$. Hence
\begin{eqnarray} \label{EQ5}
&& b = b', \quad p \overline{p} = q \overline{q} = ras rb = r'a's' r'b, \quad a < b, \; a' < b.
\end{eqnarray}
Since $q$ is a nonempty proper prefix of $p$, we have $|q| < |p|$ which yields
$|rb| = |\overline{p}|  < |\overline{q}| = |r'b|$, i.e., $|r'| > |r|$.
By Eq. (\ref{EQ5}), the word $ra$ is a prefix of $r'$, i.e., there is
a word $x$ such that $r' = rax$. Furthermore,
by Eq. (\ref{EQ5}) again, since $q$ is a nonempty proper prefix of $p$,
$\overline{p} = rb$ is a nonempty proper suffix of $\overline{q}$, i.e.,
there is a nonempty word $y$ such that $\overline{q} = yrb$.
In conclusion,
$\overline{q} = r'b = raxb = yrb$, thus $\overline{q} \ll rb$, that is,
$\overline{q}$ is not an inverse Lyndon word, a contradiction.
\enddim

\subsection{Some technical lemmas} \label{TecLem}

In this section we prove some technical results which allows us to compute
the pair $(p,s)$ in $\Pref_{bre}(w)$, for a word $w$ which is not an inverse Lyndon word.
They will be used in Section \ref{AlgoFindbre}. In detail,
Lemma \ref{basic2} is a more precise reformulation of Lemma \ref{lemintermedio}.
Lemma \ref{basic1} is preliminary to Lemma \ref{shortest} which in turn
characterizes $(p,s)$ through two simple conditions.

\begin{lemma} \label{basic2}
Let $w \in \Sigma^+$.
If $w$ has no prefix with the form $raurb$, where $r,u \in \Sigma^*$, $a,b \in \Sigma$,
with $a < b$, then $w$ is an inverse Lyndon word.
\end{lemma}
\begdim
Let $w \in \Sigma^+$ be as in the statement.
If $w$ were not an inverse Lyndon word, then, by Corollary \ref{ultimo}, there would be
$r,u \in \Sigma^*$, $a,b \in \Sigma$, with $a < b$
such that $w = raurbt$, a contradiction since $raurb$ would be
a prefix of $w$.
\enddim

\begin{lemma} \label{basic1}
Let $w, u, v \in \Sigma^+$, $r \in \Sigma^*$, and $a,b \in \Sigma$ be such that
\begin{itemize}
\item[(1)]
$w = rurv$,
\item[(2)]
$rur$ is an inverse Lyndon word,
\item[(3)]
$u$ starts with $a$, $v$ starts with $b$, with $a < b$,
\item[(4)]
$r$ is the shortest prefix of $w$ such that conditions (1)-(3) hold.
\end{itemize}
Then $w$ is not an inverse Lyndon word and $(ru, rb) \in \Pref_{bre}(w)$.
\end{lemma}
\begdim
Let $w, u, v \in \Sigma^+$, $r \in \Sigma^*$, and $a,b \in \Sigma$ be as in the statement.
By item (2) in Lemma \ref{proplexord}, $ru \ll rb$ implies
$rurb \ll rb$. Therefore, $rurb$ is not an inverse Lyndon word, hence, by
Lemma \ref{lem:inverse-Lyndon-word-prefix}
$w$ is not an inverse Lyndon word too.
Moreover, by (2), each proper nonempty prefix of $rurb$ is an inverse Lyndon word.
Thus, by Lemmas \ref{lemma2}, \ref{lemma1}, if $p$ is the longest nonempty proper prefix
of $rurb = ps$ such that
$p \ll s$ for the corresponding suffix $s$, then
$(p,s) \in \Pref_{bre}(rurb) \subseteq \Pref_{bre}(w)$.
In particular, $s$ is an inverse Lyndon word (Definition \ref{brepref}).

If $(ru, rb) \not = (p,s)$, then $|p| > |ru|$.
Moreover, since $p \ll s$, there are
$r', x, y \in \Sigma^*$, $c, d \in \Sigma$,
with $c < d$, such that $x$ starts with $c$ and $p = r'x$, $s = r'dy$.
Since any prefix of $rur$ is an inverse Lyndon word and
$r'xr'd$ is not an inverse Lyndon word, the word $rur$ is a proper prefix
of $r'xr'd$, thus, by $rurb = ps = r'xr'dy$, we have $dy = b = d$ which implies
\begin{eqnarray} \label{EQ0}
rurb & = & r'xr'b, \quad s = r'b
\end{eqnarray}
Looking at Eq.~(\ref{EQ0}), if
$|r| = |r'|$, we would have $r = r'$, hence $p = ru$,
a contradiction.
Consequently, since $r$ is the shortest prefix of $w = rurv = r'xr'v$
such that conditions (1)-(3) hold, and $r'$ satisfies conditions (1)-(3),
we have $|r'| > |r|$.
Since $u$ starts with $a$, by Eq.~(\ref{EQ0}) again,
we get $r' =rau'= u''r$.
Thus, by Proposition \ref{equazione} there are $z,t \in \Sigma^*$, $n \geq 0$, such that
$$au' = zt, \quad u''= tz,  \quad r = t(zt)^n.$$
Of course $zt \not = 1$.
If $z \not = 1$, $z$ starts with $a$ and
$s = r'b = t(zt)^nztb \ll tb$, a contradiction since $s$ is an inverse Lyndon word.
The same argument applies if $z = 1$.
In this case, $t$ starts with $a$ and
$s = r'b = t(zt)^nztb \ll b$, a contradiction since $s$ is an inverse Lyndon word.
\enddim

\begin{lemma} \label{shortest}
Let $w \in \Sigma^+$ be a word which is not an inverse Lyndon word.
Let $(p,\overline{p}) \in \Pref_{bre}(w)$.
\begin{itemize}
\item[(1)]
$p \overline{p}$ is the shortest nonempty prefix of $w$ which is not an inverse Lyndon word.
\item[(2)]
$p = rau$ and $\overline{p} = rb$, where $r,u \in \Sigma^*$, $a,b \in \Sigma$ and
$r$ is the shortest prefix of $p \overline{p}$ such that $p \overline{p} = raurb$,
with $a < b$.
\end{itemize}
\end{lemma}
\begdim
Let $w, p, \overline{p}$ be as in the statement.

(1) Let $x$ be the shortest nonempty prefix of $w$
which is not an inverse Lyndon word.
Since $x$ and $p \overline{p}$ are both prefixes of
$w$, they are comparable for the prefix order.
Since $x$ is not an inverse Lyndon word whereas any proper nonempty prefix of
$p \overline{p}$ is an inverse Lyndon word (Definition \ref{brepref}), the
word $p \overline{p}$ is a prefix of $x$, hence $|x| \geq |p \overline{p}|$.
Since $x$ is the shortest nonempty prefix of $w$
which is not an inverse Lyndon word, we have $|x| \leq |p \overline{p}|$,
hence $|x| = |p \overline{p}|$. Therefore, since $x$ and $p \overline{p}$ are
comparable for the prefix order, $x = p \overline{p}$.

(2) By Lemma \ref{lemma0}, there exists a prefix $r$
of $p \overline{p}$ such that $p \overline{p} = raurb$,
with $a < b$. The proper nonempty prefix $raur$ of
$p \overline{p}$ is an inverse Lyndon word (Definition \ref{brepref}).
Choose $r$ of minimal length.
Thus, by Lemma \ref{basic1}, $(rau, rb) \in \Pref_{bre}(p \overline{p}) \subseteq \Pref_{bre}(w)$.
\enddim

\begin{lemma} \label{shortestbis}
Let $w \in \Sigma^+$ be a word which is not an inverse Lyndon word.
The shortest nonempty prefix of $w$ which is not
an inverse Lyndon word is the shortest nonempty prefix $x$ of $w$ such
that $x = raurb$, where $r,u \in \Sigma^*$, $a,b \in \Sigma$, $a < b$.
\end{lemma}
\begdim
Let $w \in \Sigma^+$ be a word which is not an inverse Lyndon word,
let $x'$ be the shortest nonempty prefix of $w$ which is not
an inverse Lyndon word, and
let $x$ be the shortest nonempty prefix of $w$ such that
$x = raurb$, where $r,u \in \Sigma^*$, $a,b \in \Sigma$, $a < b$.
By Lemmas \ref{lemma0} and \ref{shortest},
there are $r',u' \in \Sigma^*$,
$c, d \in \Sigma$, $c < d$, such that $x' = r'cu'r'd$.
The words $x,x'$ are both
prefixes of $w$, thus they are comparable for the prefix order.
If they were different, by the hypothesis on $|x|$, $x$ would be
a proper nonempty prefix of $x'$. This fact contradicts the hypothesis on $|x'|$
because $x$ is not an inverse Lyndon word (Corollary \ref{ultimo}).
\enddim

\begin{example} \label{shortestnecessario}
{\rm Let $\Sigma = \{a,b,c,d \}$ with $a < b < c < d$,
let $w = cbabcbad \in \Sigma^+$. The word $w$ is not an inverse Lyndon word but
any nonempty proper prefix of $w$ is an inverse Lyndon word.
By Lemma \ref{shortest}, item (1), we have
$w = p \overline{p}$, with $(p,\overline{p}) \in \Pref_{bre}(w)$.
We have $cbab \ll cbad$ but $cba$ is not the shortest
prefix $r$ of $w = p \overline{p}$ satisfying item (2) in lemma \ref{shortest}.
Notice that $cbad$ is not an inverse Lyndon word.
Since $cbabcba \ll d$,
the shortest prefix $r$ of $w = p \overline{p}$ satisfying item (2) in lemma \ref{shortest}
is the empty word. Consequently, $(p,\overline{p}) = (cbabcba, d)$.}
\end{example}

\begin{example}  \label{Ebre1}
{\rm Let $\Sigma = \{a,b,c,d \}$ with $a < b < c < d$, let
$v = dabdabdadac$. We can check that
$dabdabdad$ is the shortest nonempty prefix of
$v$ which is not an inverse Lyndon word, hence
$dabdabdad = p \overline{p}$, with $(p,\overline{p}) \in \Pref_{bre}(v)$.
The shortest prefix $r$ of $p \overline{p}$ satisfying item (2) in lemma \ref{shortest}
is $da$, thus
$(p,\overline{p}) = (dabdab,dad) \in \Pref_{bre}(v)$.
As another example, consider $w=dabadabdabdadac$. We can check that
$dabadabd$ is the shortest nonempty prefix of
$w$ which is not an inverse Lyndon word, hence
$dabadabd = q \overline{q}$, with $(q,\overline{q}) \in \Pref_{bre}(w)$.
The shortest prefix $r$ of $q \overline{q}$ satisfying item (2) in lemma \ref{shortest}
is $dab$, thus
$(q,\overline{q}) = (daba, dabd) \in \Pref_{bre}(w)$.}
\end{example}

\begin{example} \label{Ebre2}
{\rm Let $\Sigma = \{a, b, c \}$ with $a < b < c$,
let $v =cbabacbac$.
We can check that $v = cbabacbac$
is the shortest nonempty prefix of
$v$ which is not an inverse Lyndon word, hence
$v = cbabacbac = p \overline{p}$, with $(p,\overline{p}) \in \Pref_{bre}(v)$.
The shortest prefix $r$ of $p \overline{p}$ satisfying item (2) in lemma \ref{shortest}
is $cba$, thus
$(p,\overline{p}) = (cbaba, cbac) \in \Pref_{bre}(v)$.
As another example, consider $w = cbabacaacbabacbac$.
We can check that
$cbabacaacbabacb$ is the shortest nonempty prefix of
$w$ which is not an inverse Lyndon word, hence
$cbabacaacbabacb = q \overline{q}$, with $(q,\overline{q}) \in \Pref_{bre}(w)$.
The shortest prefix $r$ of $q \overline{q}$ satisfying item (2) in lemma \ref{shortest}
is $cbabac$, thus
$(q,\overline{q}) = (cbabacaa, cbabacb) \in \Pref_{bre}(w)$.}
\end{example}

\subsection{A canonical inverse Lyndon factorization: $\ICFL(w)$} \label{icfl}

We give below the recursive definition of the canonical inverse Lyndon factorization
$\ICFL(w)$.

\begin{definition} \label{def:ICFL}
Let $w \in \Sigma^+$.

\medskip\noindent
{\rm (Basis Step)}
If $w$ is an inverse Lyndon word,
then $\ICFL(w) = (w)$.

\medskip\noindent
{\rm (Recursive Step)}
If $w$ is not an inverse Lyndon word,
let $(p,\overline{p}) \in \Pref_{bre}(w)$ and let
$v \in \Sigma^*$ such that $w = pv$.
Let $\ICFL(v) = (m'_1, \ldots, m'_{k})$ and let
$r \in \Sigma^*$ and $a,b \in \Sigma$ such that $p = rax$, $\overline{p} = rb$ with $a < b$.
$$\ICFL(w) = \begin{cases} (p, \ICFL(v)) & \mbox{ if } \overline{p} = rb \leq_{p} m'_1 \\
(p m'_1, m'_2, \ldots, m'_{k}) & \mbox{ if } m'_1 \leq_{p} r \end{cases}$$
\end{definition}

\begin{remark}
{\rm With the same notations as in the recursive step of Definition \ref{def:ICFL},
we notice that $rb$ and $m'_1$ are both prefixes of the same word $v$.
Thus they are comparable for the prefix order, that is, either $rb \leq_{p} m'_1$
or $m'_1 \leq_{p} r$.}
\end{remark}

\begin{example} \label{Ex1}
{\rm Let $\Sigma = \{a, b, c \}$ with $a < b < c$, let
$v = cbabacbac$. Let us compute $\ICFL(v)$.
As showed in Example \ref{Ebre2}, we have
$(p, \bar{p}) = (cbaba, cbac) \in \Pref_{bre}(v)$.
Therefore, $cbaba, cbac$ are both inverse Lyndon words and
we have
$\ICFL(cbaba) = (cbaba)$,
$\ICFL(\bar{p}) = \ICFL(cbac) = (m') = (cbac)$.
Since
$\bar{p} = cbac \leq_{p} m'$, we are in the first case
of Definition \ref{def:ICFL}, hence
$\ICFL(v) = \ICFL(cbabacbac) = (p, \ICFL(\bar{p}))
= (cbaba, cbac) = (m'_1,m'_2)$.
Now, let $w = cbabacaacbabacbac$ and let us compute
$\ICFL(w)$.
In Example \ref{Ebre2} we showed that
$(q, \bar{q}) = (cbabacaa, cbabacb) \in \Pref_{bre}(w)$.
Thus, $w = qv$, where $v = cbabacbac$ is the above considered
word and $\bar{q} = r'b$, where $r' = cbabac$.
Since $m'_1 = cbaba \leq_p cbabac$,
we are in the second case of Definition \ref{def:ICFL}, therefore
$\ICFL(w) = (q m'_1, m'_2) = (cbabacaacbaba, cbac)$.}
\end{example}

\begin{example} \label{Ex2}
{\rm Let $\Sigma = \{a,b,c,d \}$ with $a < b < c < d$, let
$v=dabdabdadac$. Let us compute $\ICFL(v)$.
As showed in Example \ref{Ebre1}, we have
$(p, \bar{p}) = (dabdab, dad) \in \Pref_{bre}(v)$.
Hence, $v = pv'$, where $v' = dadac = \overline{p}ac$.
On the other hand, we can easily see
that $dadac$ is an inverse Lyndon word, hence $\ICFL(dadac)=(dadac)$.
Since $\overline{p}=dad \leq_p dadac$, by
Definition \ref{def:ICFL} (first case),
$\ICFL(v) = (dabdab, dadac)$.
Now, let us compute $\ICFL(w)$, where $w=dabadabdabdadac$.
In Example \ref{Ebre2} we showed that
$(q, \bar{q}) = (daba, dabd) \in \Pref_{bre}(w)$.
Thus, $w = qv$, where $v$ is the above considered
word. Since $\overline{q}=dabd \leq_p dabdab$, by
Definition \ref{def:ICFL} (first case),
$\ICFL(w) = (daba, dabdab, dadac)$.
In Example \ref{nonunique} we showed
two different inverse Lyndon factorizations of $w' = dabdadacddbdc$.
Both are different from
$\ICFL(w') = (dab, dadac, ddbdc)$.}
\end{example}

As proved below, $\ICFL(w)$ is uniquely determined.

\begin{proposition} \label{unique}
For any word $w \in \Sigma^+$, there is a unique sequence $(m_1, \ldots, m_k)$ of words
over $\Sigma$ such that $\ICFL(w) = (m_1, \ldots, m_k)$.
\end{proposition}
\begdim
The proof is by induction on $|w|$. If $w$ is a letter, then the statement is proved
since $w$ is an inverse Lyndon word and $\ICFL(w) = (w)$ (Definition \ref{def:ICFL}).
Thus, assume $|w| > 1$.
If $w$ is an inverse Lyndon word we have done since, by Definition \ref{def:ICFL}, $\ICFL(w) = (w)$.
Otherwise, $w$ is not an inverse Lyndon word and there is a unique pair
$(p,\overline{p})$ in $\Pref_{bre}(w)$ (Proposition \ref{unica}). Let
$v \in \Sigma^+$ such that $w = pv$.
Since $|v| < |w|$, by induction hypothesis there is a unique sequence
$(m'_1, \ldots, m'_{k'})$ of words such that
$\ICFL(v) = (m'_1, \ldots, m'_{k'})$. Let
$r \in \Sigma^*$ and $a,b \in \Sigma$ such that $p = rax$, $\overline{p} = rb$ with $a < b$.
By the recursive step of Definition \ref{def:ICFL}, we have
$$\ICFL(w) = \begin{cases} (p, \ICFL(v)) & \mbox{ if } rb \leq_{p} m'_1 \\
(p m'_1, m'_2, \ldots, m'_{k'}) & \mbox{ if } m'_1 \leq_{p} r \end{cases}$$
In both cases, by the above arguments, the sequence $\ICFL(w)$ is uniquely determined.
\enddim

As a main result, we now prove that $\ICFL(w)$ is an inverse Lyndon factorization of $w$.

\begin{lemma}\label{lem:ICFL-factorization-1}
For any $w \in \Sigma^+$, the sequence $\ICFL(w) = (m_1, \ldots, m_k)$ is an inverse Lyndon
factorization of $w$, that is, $w = m_1 \cdots m_k$,
$m_1 \ll  \ldots \ll m_k$ and each $m_i$ is an inverse Lyndon word.
\end{lemma}
\begdim
The proof is by induction on $|w|$. If $w$ is a letter, then the statement is proved
since $w$ is an inverse Lyndon word and $\ICFL(w)= (w)$ (Definition \ref{def:ICFL}).
Thus, assume $|w| > 1$.
If $k = 1$ we have done since, by Definition \ref{def:ICFL},
$w$ is an inverse Lyndon word and $\ICFL(w) = (w)$.
Otherwise, $w$ is not an inverse Lyndon word. Let $(p,\overline{p}) \in \Pref_{bre}(w)$ and let
$v \in \Sigma^+$ such that $w = pv$.
Let $\ICFL(v) = (m'_1, \ldots, m'_{k'})$ and let
$r, x \in \Sigma^*$ and $a,b \in \Sigma$ such that $p = rax$, $\overline{p} = rb$ with $a < b$.
By the recursive step of Definition \ref{def:ICFL}, we have
$$\ICFL(w) = \begin{cases} (p, \ICFL(v)) & \mbox{ if } rb \leq_{p} m'_1 \\
(p m'_1, m'_2, \ldots, m'_{k'}) & \mbox{ if } m'_1 \leq_{p} r \end{cases}$$
Moreover, since $|v| < |w|$, by induction hypothesis, $v = m'_1 \cdots m'_{k'}$,
$m'_1 \ll  \ldots \ll m'_{k'}$ and each $m'_i$ is an inverse Lyndon word.
Therefore, $w = pv = p m'_1 \cdots m'_{k'}$.

If $rb \leq_{p} m'_1$, then there is $z \in \Sigma^*$ such that
$m'_1 = rbz$ and
$\ICFL(w) = (p, m'_1, \ldots, m'_{k'})$. By Definition \ref{brepref},
$p$ is an inverse Lyndon word and so are all the words in $\ICFL(w)$.
Furthermore, $p \ll m'_1 \ll  \ldots \ll m'_{k'}$ and the proof is ended.
Otherwise, there is $z \in \Sigma^*$ such that
$r = m'_1z$, hence $p = rax = m'_1z ax$ and $\ICFL(w) = (p m'_1, \ldots, m'_{k'})$.
The word $p m'_1$ is a proper prefix of $p \overline{p}$, thus, by Definition \ref{brepref},
$p m'_1$ is an inverse Lyndon word and so are all words in $\ICFL(w)$.
Finally, by $m'_1 \ll m'_2 \ldots \ll m'_{k'}$
we have $p m'_1 = m'_1z ax m'_1 \ll m'_2 \ldots \ll m'_{k'}$ (Lemma \ref{proplexord}).
\enddim

We end this section with the following result showing that inverse
Lyndon factorizations of Lyndon words are not trivial.

\begin{proposition} \label{main1}
Any word $w \in \Sigma^+$ has an inverse Lyndon factorization $(m_1, \ldots , m_k)$.
Moreover, if $w$ is a Lyndon word which is not a letter, then $k> 1$.
\end{proposition}
\begdim
The first part of the statement follows by
Proposition \ref{unique} and Lemma \ref{lem:ICFL-factorization-1}.
Assume
that $w$ is a Lyndon word which is not a letter
and let $(m_1, \ldots , m_k)$ be one of its inverse Lyndon factorizations.
By Proposition \ref{P2}, there is a proper
nonempty suffix $v$ of $w$ such that $w \prec v$. Hence,
$w$ is not an inverse Lyndon word, which yields $w \not = m_1$
and consequently $k > 1$.
\enddim

Of course a converse of Proposition \ref{main1} can also be stated.

\begin{proposition} \label{viceversa}
Let $w \in \Sigma^+$ and let $\CFL(w) = (\ell_1, \ldots, \ell_h)$.
If $w$ is an inverse Lyndon word which is not a letter, then $h > 1$.
\end{proposition}
\begdim
Let $w$ be an inverse Lyndon word which is not a letter and
let $\CFL(w) = (\ell_1, \ldots, \ell_h)$.
By Definition \ref{inverse-Lyndon-word},
there is a proper
nonempty suffix $v$ of $w$ such that $v \prec w$.
Hence, by Proposition \ref{P2},
$w$ is not a Lyndon word, which yields $w \not = \ell_1$
and consequently $h > 1$.

\section{An algorithm for finding the bounded right extension} \label{AlgoFindbre}

In section \ref{AlgoICFL}
we will give a linear time recursive algorithm, called \textbf{Compute-ICFL}, that
computes $\ICFL(w)$, for a given nonempty word $w$.
By Definition \ref{def:ICFL}, we know that the computation of $\ICFL(w)$, when $w$ is
not an inverse Lyndon word, is based on that of the pair $(p,\overline{p}) \in \Pref_{bre}(w)$.
In this section we give algorithms to compute the above pair $(p,\overline{p})$.
By Lemmas \ref{shortest}, \ref{shortestbis}, we are faced with the problems of
\begin{itemize}
\item[(1)]
stating whether $w$ is an inverse Lyndon word;
\item[(2)]
if not, finding the shortest prefix $x$ of $w$ such
that $x = raurb$, where $r,u \in \Sigma^*$, $a,b \in \Sigma$, $a < b$.
Therefore, $x = p \bar{p}$.
\item[(3)]
Finding the shortest $r$ such that $x = raurb$, with $r, u, a, b$ as in (2).
Thus $p = rau$, $\bar{p} = rb$.
\end{itemize}
The first two tasks are carried out by algorithm \textbf{Find-prefix},
the third is performed by algorithm \textbf{Find-bre}.
Algorithm \textbf{Find-prefix} is very similar to Duval's algorithm in \cite{duval} which computes
the longest Lyndon prefix of a given string $w$. This is not surprising because
\textbf{Find-prefix} computes the longest prefix $raur$ of $w$ which is an inverse Lyndon word.
Algorithm \textbf{Compute-ICFL} uses algorithm \textbf{Find-bre} as a subroutine.
In turn, the latter uses the output of algorithm \textbf{Find-prefix}.
Algorithm \textbf{Find-bre} also calls a procedure to compute the well known {\it failure function}
(named the  {\it prefix function} in \cite{CLR}) of the
Knuth-Morris-Pratt matching algorithm \cite{KMP}.

Firstly, we give a high-level description of algorithm \textbf{Find-prefix}
followed by its pseudocode (Section \ref{DescFindpre}) and we prove its correctness through
some loop invariants (Section \ref{CorrectFindpre}).
Next, in Section \ref{failure}, we recall the definition of the failure function
and we prove some results concerning this function which will be useful later.
Finally, we give a
high-level description of algorithm \textbf{Find-bre}
followed by its pseudocode (Section \ref{DescFindbre}) and
we end the section with the proof of its correctness through
some loop invariants (Section \ref{CorrectFindbre}).

\subsection{Description of Find-prefix} \label{DescFindpre}

Let us describe the high-level structure of algorithm
\textbf{Find-prefix}. As already said, given $w$, the algorithm looks for the shortest prefix
$raurb$ of $w$, where $r,u \in \Sigma^*$, $a,b \in \Sigma$, $a < b$ and
$raur$ is an inverse Lyndon word. This is equivalent to find, if it exists,
a nonempty word $r$ which is a prefix
of $w$ and a prefix of a suffix of $w$, and these two occurrences of $r$ are followed
by $a,b$ as above.

As usual, the word is represented as an array $w[1..n]$ containing the sequence of
the letters in $w$, with $n =|w|$.
The algorithm uses two indices $i,j$ to scan the word $w$. Initially, these indices
denote the position of the first letter of a candidate common prefix $r$.

A while-loop is used
to compare the two letters $w[i]$ and $w[j]$. While $j$ is incremented at each iteration,
$i$ is incremented only when $w[i] = w[j]$. Notice that the algorithm does not test
if eventually the two occurrences of $r$ overlap.
Lemma \ref{basic3} allows us to avoid this test.

If $w[i] > w[j]$, then the algorithm resets $i$ to the first
position of $w$ and examines a new candidate common prefix $r$, whose
position of the first letter in the second
occurrence of $r$ is indicated by the value of $j$ at this time.

The loop condition is false when $w[i] < w[j]$ or when $j$ denotes the last letter of $w$.
In the first case, our search has been successful and the algorithm returns $raurb$.
If $j$ denotes the last letter of $w$ (and $w[i] \geq w[j]$), then $w$ has no prefix
$raurb$, with $r,u,a,b$ as above, and the algorithm returns $w \$$, where $\$$ is
a letter such that $\$ \not \in \Sigma$. In this case, by Lemma
\ref{basic2}, we know that $w$ is an inverse Lyndon word.

In conclusion, \textbf{Find-prefix} allows us to state whether $w$ is an inverse Lyndon word
or not and, in the latter case, it finds the prefix $x$ of $w$ such that
$x = p \bar{p}$, with $(p, \bar{p}) \in \Pref_{bre}(w)$.
Algorithm \ref{alg2} describes the procedure \textbf{Find-prefix}.
It is understood that the empty array, namely $w[j+1..n]$ with $j=n$,
represents the empty word.

\begin{algorithm}[htb!]
\SetKwInOut{PRE}{Input}
\SetKwInOut{POST}{Output}
\PRE{A string $w$}
\POST{A pair of strings $(x,y)$, where $xy =w$, and moreover
$x=w\$$, $y = 1$ if $w$ is an inverse Lyndon word,
$x = p \bar{p}$, with $(p, \bar{p}) \in \Pref_{bre}(w)$, otherwise.}

\If{$|w| = 1$ }{
  \Return $(w\$,1)$\;
 }
 $i \gets 1$\;
 $j \gets 2$\;

\While{$j < |w|$ and $w[j] \leq  w[i]$}
 {
 \If{$w[j] < w[i]$}
{
$i \gets 1$\;
}
\Else{$i \gets i + 1$}
 $j \gets j +1$\;
 }
 \If{$j = |w|$}
{

\If{$w[j] \leq w[i]$}
  {\Return $(w\$$,$1)$\;}
}
  \Return $(w[1..j]$, $w[j+1..|w|])$ \;

\caption{Find-prefix}
\label{alg2}
\end{algorithm}

We notice that we can reach line (11) in two different cases: either when $x = w$
or when $w$ is an inverse Lyndon word.
We distinguish these two cases: in the former case the output is $(w ,1)$
and in the latter case the output is $(w\$ ,1)$ (see example below).

\begin{example} \label{uno}
{\rm Let $\Sigma = \{a, b, c \}$ with $a < b < c$.
Algorithm \textbf{Find-prefix}$(bac)$ returns $(bac,1)$. Indeed, $|w| = 3$ and
$w[2] <  w[1]$ in the first iteration of the while-loop. Then, $i = 1$ (line $(7)$),
$j = 3$ (line $(10)$), and we reach line $(11)$ with $j = 3 = |w|$.
Since $w[j] = w[3] = c > w[1] = w[i]$,
the algorithm returns  $(w[1..j], w[j+1..|w|]) = (bac,1)$.

Consider now the inverse Lyndon word $w=bab$.
Algorithm \textbf{Find-prefix}$(bab)$ returns $(bab\$,1)$. Indeed, $|w|=3$ and, as before,
$w[2] <  w[1]$ in the first iteration of the while-loop. Then, $i = 1$ (line $(7)$),
$j = 3$ (line $(10)$), and we reach line $(11)$ with $j = 3 = |w|$.
Since $w[j] = w[3] = b = w[1] = w[i]$,
the algorithm returns $(bab\$ , 1)$.
The same argument applies when $w=baa$.}
\end{example}

\begin{example} \label{t}
{\rm Let $\Sigma = \{a, b \}$ with $a < b$.
Algorithm \textbf{Find-prefix}$(bbabbabbb)$ outputs $(bbabbabbb,1)$. }
\end{example}

\subsection{Correctness of Find-prefix} \label{CorrectFindpre}

In this section we prove that \textbf{Find-prefix} does what it is claimed to do.
To begin, notice that each time around the while-loop of lines (5) to (10),
$j$ is increased by 1 at line (10). Therefore, when $j$ becomes equal to $|w|$,
if we do not break out of the while-loop earlier, the loop condition $j < |w|$
will be false and the loop will terminate.
In order to prove that \textbf{Find-prefix} does what it is intended to do,
we define the following loop-invariant statement, where we use $k$
to stand for one of the values that the variable
$j$ assumes and $h_k$ for the corresponding
variable $i$, as we go around the loop.
Finally $h'_k$ depends on $h_k$ and $k$.

\begin{quote}
$S(k)$:
If we reach the loop test ``$j < |w|$ and $w[j] \leq w[i]$''
with
the variable $i$ having the value
$h_k$ and
the variable $j$ having the value $k$, then
\begin{itemize}
\item[(a)]
$1 \leq h_k < k$. Set $h'_k = k - h_k + 1$. We have $1 < h'_k \leq |w|$ and the following conditions
holds
\begin{itemize}
\item[(a1)]
$w[1..h_k-1] = w[h'_k..k-1]$, that is, $w[1..h_k-1]$ is a proper
prefix of $w[1..k-1]$ and $w[1..h_k-1]$ is also a suffix of $w[1..k-1]$.
\item[(a2)]
For any $t'$, $1 < t' < h'_k$, $w[t' ..k-1]$ is not a prefix of $w[1..k-1]$.
\end{itemize}
\item[(b)]
$w[1..k-1]$ has no prefix with the form $raurb$, $r,u \in \Sigma^*$, $a,b \in \Sigma$, $a < b$.
Therefore, by Lemma \ref{basic2},
$w[1..k-1]$ is an inverse Lyndon word.
\end{itemize}
\end{quote}

\begin{remark} \label{nonempty}
{\rm Notice that  $w[t' ..k-1]$
is nonempty, for any  $t'$, $1 < t' < h'_k$.
Indeed, $h'_k = k - h_k + 1 \leq k$. Thus, since $t' \leq h'_k -1 \leq k-1$,
we have $|w[t' ..k-1]| \geq |w[h'_k -1 ..k-1]| \geq |w[k-1..k-1]| = |w[k-1]| = 1$.}
\end{remark}

Loosely speaking, at the beginning of each iteration of the loop of
lines (5)-(10), indices $i,j$ store the end of a candidate common prefix $r$ (item (a1))
and we do not yet find a prefix with the required form in the examined part
of the array (items (a2), (b)).
Note that the examined part may be a single letter and $r$ could be the empty word.

We need to show that the above loop invariant is true prior to the first iteration
and that each iteration of the loop maintains the invariant.
This will be done in Proposition \ref{InizMaint}, where we prove $S(k)$ by (complete) induction on $k$.
Then, we show that the invariant provides a useful property to prove correctness
when the loop terminates in Proposition \ref{termination}.
Even if $S(k)$ clearly holds for $k = 0,1$, we prove it for $k \geq 2$.

\begin{proposition} \label{InizMaint}
For any $k \geq 2$, $S(k)$ is true.
\end{proposition}
\begdim
{\bf (Basis)} Let us prove that $S(2)$ is true. We reach the test with
$j$ having value $2$ only when we enter the loop from the outside.
Prior to the loop, lines $(3)$ and $(4)$ set $i$ to 1 and $j$ to 2.
Therefore $h_k = 1 < k = 2$ and $h'_k = k - h_k  + 1 = 2$, hence
$1 < h'_k \leq |w|$.
Part (a1) of $S(2)$ is true. Indeed $w[1..1-1] = w[2..2-1]$
holds since there are no elements in both the descriptions.
Part (a2) of $S(2)$ is also true since there are no integers $t'$ such that $1 < t' < 2$.
Similarly, since $w[1..2-1] = w[1]$, part (b) is true.

\medskip
\noindent
{\bf (Induction)} We suppose that $S(k)$ is true and prove that $S(k+1)$ is true.
We may assume $k < |w|$ and $w[k] \leq w[h_k]$ (otherwise we break out of the while-loop
when $j$ has the value $k$ or earlier
and $S(k + 1)$ is clearly true, since it is a conditional expression with a false antecedent).
Moreover, by using the inductive hypothesis (part (a)), $1 \leq h_k < k$,
$w[1..h_k-1] = w[h'_k ..k-1]$, where $h'_k = k - h_k + 1 > 1$ and
$w[t' ..k-1]$ is not a prefix of $w[1..k-1]$, $1 < t' < h'_k$.

To prove $S(k+1)$, we consider what happens when we execute the body of the while-loop
with $j$ having the value $k$ and $i$ having the value $h_k$.
Moreover, to prove part (a) of $S(k+1)$, we distinguish two cases:
\begin{itemize}
\item[(1)]
$w[k] < w[h_k]$,
\item[(2)]
$w[k] = w[h_k]$.
\end{itemize}
(Case (1)) If $w[k] < w[h_k]$, then line $(7)$ set $i$ to 1.
Thus, when we reach the loop test with $j$ having the value $k+1$, the variable $i$
has the value $h_{k+1} = 1$. Consequently, $h'_{k+1} = k + 1 - h_{k+1} + 1 = k + 1$
which implies $1 < h'_{k+1} = k + 1 \leq |w|$ and
$w[1..h_{k+1} - 1] = w[1..0] = w[k+1..k]$. Therefore, part (a1) is proved in this case.
Let us prove (a2). Let $t'$ be any integer such that $1 < t' < h'_{k+1} = k +1$. If
$w[t' ..k]$ were a prefix of $w[1..k]$, then
there would exist $t$ such that $w[1..t] = w[t' ..k]$, with $t \geq 1$
(see Remark \ref{nonempty}).
Moreover, $w[t' ..k-1]$ would be a prefix of $w[1..k-1]$ with $t' >1$ thus,
by inductive hypothesis (part (a2)), we would have $t' \geq h'_k$.
Consequently, $|w[1..t]| = |w[t' ..k]| \leq |w[h'_k ..k]| = |w[h'_k ..k -1]| + 1
= |w[1..h_k-1]| + 1$
which implies $t \leq h_k$.
Notice that $w[t] = w[k] < w[h_k]$, hence $t \leq h_k - 1$.

Since $t \leq h_k - 1$, the word $w[1..t-1] = w[t' ..k-1]$ is a proper prefix of
$w[1..h_k -1]$ and, since $t' \geq h'_k$, the same word $w[1..t-1] = w[t' ..k-1]$ is a suffix of
$w[h'_k ..k-1] = w[1..h_k-1]$.
Notice that each time around the while-loop, $j$ is increased by $1$ at line $(10)$
and $2 \leq h_k + 1 \leq k$, by inductive hypothesis (part (a)).
Therefore, when we reach the loop test with $j$ having the value $h_k + 1$, the word
$w[1..h_k]$ has a prefix with the form $raurb$, $r,u \in \Sigma^*$, $a,b \in \Sigma$, $a < b$.
Indeed, $w[1..h_k] = raurb$, where $a = w[t] < w[h_k] = b$ and $r = w[1..t-1]$ if
$t = 1$ or if
the two occurrences of $w[1..t-1]$ (as a prefix and a suffix) in $w[1..h_k-1]$
do not overlap, otherwise it is its prefix given by Lemma \ref{basic3} applied
to $w[1..t-1]$. Hence part (b) of $S(h_k + 1)$ is not true, a contradiction.

(Case (2)) Assume that $w[k] = w[h_k]$.
Thus, by $w[1..h_k-1] = w[h'_k ..k-1]$, we also have
$w[1..h_k] = w[h'_k ..k]$.
Therefore, when we reach the loop test with $j$ having the
value $k+1$, the variable $i$ has value $h_{k+1} = h_k +1$.
Consequently,
$h'_{k+1} = k + 1 - h_{k+1} + 1 = k - h_k + 1 = h'_k$ and
$w[1..h_{k+1} - 1]= w[1..h_k] = w[h'_k..k] = w[h'_{k+1}..k]$,
with $1 < h'_{k+1} = h'_k \leq |w|$. Thus part (a1) is proved in case (2).
Furthermore, by inductive hypothesis (part (a2)),
for any $t'$, $1 < t' < h'_k = h'_{k+1}$, the word $w[t' ..k-1]$ is not a prefix of $w[1..k-1]$,
hence $w[t' ..k]$ is not a prefix of $w[1..k]$ and part (a2) holds.

To prove part (b) of $S(k+1)$, assume on the contrary that
$w[1..k]$ has a prefix with the form $raurb$,
$r,u \in \Sigma^*$, $a,b \in \Sigma$, $a < b$.
Thus, by using the inductive hypothesis
(part (b)), we would have $w[1..k] = raurb$, that is $w[k] = b$.
Moreover, there would exist $t, t'$,
with $1 \leq t = |r| + 1 < k$ and $1 < t' \leq k-1$, such that $w[t] = a$
and $r = w[1..t-1] = w[t'..k-1]$.
Hence,
\begin{equation} \label{EQ9}
w[t] = a < b = w[k] \leq w[h_k].
\end{equation}
Since $t < k$, the word $w[1..t-1] = w[t'..k-1]$ is a prefix of $w[1..k-1]$.
Therefore, by using the inductive hypothesis, part (a2), we have $t' \geq h'_k$.
Now we apply the same argument as above.
By $t' \geq h'_k$, we have $|w[1..t -1]| = |w[t'..k -1]| \leq |w[h'_k..k -1]|
= |w[1..h_k-1]|$ which implies $t - 1 \leq h_k - 1$.
Moreover $t - 1 < h_k - 1$, by Eq. (\ref{EQ9}).
In turn, since $t - 1 < h_k - 1$, the word $r = w[1..t-1] = w[t'..k-1]$ is a proper prefix of
$w[1..h_k -1]$ and, since $t' \geq h'_k$, the same word $w[1..t-1] = w[t'..k-1]$ is a suffix of
$w[h'_k..k-1] = w[1..h_k-1]$.
As already noticed, each time around the while-loop, $j$ is increased by $1$ at line $(10)$
and $1 \leq t < h_k < k$ (see above and the inductive hypothesis, part (a)).
Therefore, when we reach the loop test with $j$ having the value $h_k + 1$, we have
$w[1..h_k] = r'au'r'b'$, where $u' \in \Sigma^*$, $b' \in \Sigma$,
$a = w[t] < w[h_k] = b'$ (Eq. (\ref{EQ9})) and $r' = r$ if $r = 1$ or if
the two occurrences of $r$ (as a prefix and a suffix) in $w[1..h_k-1]$
do not overlap, otherwise it is its prefix $r'$ given by Lemma \ref{basic3} applied
to $r$. Hence part (b) of $S(h_k + 1)$ is not true, a contradiction.

Of course, when we reach the loop test with $j$ having the
value $k+1$, the variable $i$ has either value $h_{k+1} = 1$ or
$h_{k+1} = h_k + 1$. We always have $k + 1 > 1$ and, if $1 \leq h_k < k$, then $1 \leq h_k + 1 < k +1$.
Thus, we have proved the inductive step.
\enddim

\begin{proposition} \label{termination}
Algorithm \textbf{Find-prefix} allows us to state whether $w$ is an inverse Lyndon word
or not and, in the latter case, it finds the shortest prefix $x$ of $w$ such that
$x = raurb$ where $r,u \in \Sigma^*$, $a, b \in \Sigma$, $a < b$.
More precisely,
if $w$ is an inverse Lyndon word, then \textbf{Find-prefix} outputs $(w\$,1)$;
if $w$ is not an inverse Lyndon word, then \textbf{Find-prefix} outputs
$(p\bar{p},y)$, where $w = p\bar{p}y$ and $(p, \bar{p}) \in \Pref_{bre}(w)$.
\end{proposition}
\begdim
Let $n = |w|$.
We have already shown that the while-loop will terminate.
Evidently it terminates in two cases:
when $w[i] < w[j]$, $j \leq n$, or when $j = n$ and $w[i] \geq w[n]$.
Moreover, when we reach line $(11)$, statement $S(j)$ must hold.
In particular, $w[1..i-1] = w[h'_j ..j-1]$,
where $1 < h'_j = j - i + 1 \leq n$, $1 \leq i < j$.

If $w[i] < w[j]$, the word $w[1..j]$ is a prefix of $w = w[1..n]$ which has the form
$raurb$ where $r,u \in \Sigma^*$, $a = w[i] < w[j] = b$
and $r = w[1..i-1] = w[h'_j..j-1]$, if $r = 1$ or if
the two occurrences of $r$ (as a prefix and a suffix) in $w[1..j-1]$
do not overlap, otherwise it is its prefix $r'$ given by Lemma \ref{basic3} applied
to $r$. By Corollary \ref{ultimo}, $w[1..j] = raurb$ is not an inverse Lyndon word.
On the contrary, by part (b) of $S(j)$, $raur = w[1..j-1]$ is an inverse Lyndon word. Thus,
$raurb$ is the shortest prefix of $w$ which is not an inverse Lyndon word, hence,
by Lemma \ref{shortest}, $raurb = w[1..j] = p\bar{p}$, where
$(p, \bar{p}) \in \Pref_{bre}(w)$. Finally, \textbf{Find-prefix} outputs
$(w[1..j], w[j+1..|w|]) = (p\bar{p},y)$, where $w = p\bar{p}y$ (line (14)).

If $j =n$ and $w[i] \geq w[n]$, then we prove below that
$w$ has no prefix with the form $raurb$, $r,u \in \Sigma^*$, $a,b \in \Sigma$, $a < b$.
As a consequence, $w$ is an inverse Lyndon word (Lemma \ref{basic2})
and \textbf{Find-prefix} outputs $(w\$,1)$ (line (13)).
If $w$ had a prefix with the form $raurb$, $r,u \in \Sigma^*$, $a,b \in \Sigma$, $a < b$,
then, by part (b) of $S(j) = S(n)$,
we would have $w = raurb$, that is $w[n] = b$.
Moreover, there would exist $t,t'$, with $1 \leq t = |r| + 1< n$,
$1 < t' \leq n$, such that
$r = w[1..t-1] = w[t'..n-1]$ would be a prefix and a suffix of
$w[1..n-1]$ and
\begin{equation} \label{EQ10}
w[t] = a < b = w[n] \leq w[i].
\end{equation}
By using part (a2) of $S(n)$,  we would have
$t' \geq h'_n$, hence
$|r| = |w[1..t -1]| = |w[t' ..n -1]| \leq |w[h'_n ..n -1]|
= |w[1..i-1]|$, which implies $t - 1 \leq i - 1$ and finally
$t - 1 < i - 1$, by Eq. (\ref{EQ10}).
In turn, since $t - 1 < i - 1$, the word $r = w[1..t-1] = w[t' ..n-1]$ would be a proper prefix of
$w[1..i -1]$ and, since $t' \geq h'_n$, the same word $r = w[1..t-1] = w[t'..n-1]$
would be a suffix of
$w[h'_n..n-1] = w[1..i-1]$.
In conclusion, $w[1..i] = r'aur'b'$, where $b' \in \Sigma$,
$a = w[t] < b = w[n] \leq w[i] = b'$, and $r' = r$ if $r = 1$ or if
the two occurrences of $r$ (as a prefix and a suffix) in $w[1..i-1]$
do not overlap, otherwise it is its prefix $r'$ given by Lemma \ref{basic3} applied
to $r$. Hence part (b) of $S(i + 1)$ would be false, with $i + 1 \leq j = n$, a contradiction.
\enddim

\subsection{The failure function} \label{failure}

We recall that, given a word, represented by the array $w[1..n]$,
the failure function for $w$, is the function
$f : \{1, \dots , n \} \rightarrow \{0, \ldots , n-1 \}$
such that
$$f(i) = \max\{k ~|~ k < i \mbox{ and } w[1..k] \mbox{ is a suffix of } w[1..i] \}.$$
That is, $f(i)$ is the length of the longest prefix of $w[1..n]$ which is a proper
suffix of $w[1..i]$. Thus, by Lemma \ref{basic4}, $f(i)$ is the length of the longest
proper prefix of $w[1..i]$ which is a suffix of $w[1..i]$.
It is known that by iterating the failure function $f$, we can
enumerate all the prefixes $w[1..k]$ that are suffixes of a given
prefix $w[1..i]$ (Lemma \ref{iterationlemma}). A slightly more precise version of this result
is given in Lemma \ref{prefix-property}.

\begin{definition} \label{iteratedfailure}
For a positive integer $i$, we set
\begin{eqnarray*}
f^{(0)}(i) &=& i \\
f^{(\ell)}(i) &=& f (f^{(\ell-1)}(i)), \mbox{ for } \ell \geq 1 \\
f^*(i) &=& \{i, f(i), f^{(2)}(i), \ldots , f^{(m)}(i) \},
\end{eqnarray*}
where it is understood that the sequence in $f^*(i)$ stops when $f^{(m)}(i) = 0$
is reached.
\end{definition}

\begin{lemma} \label{comm}
For any positive integer $i$ and $\ell \geq 1$
$$f^{(\ell)}(i) = f (f^{(\ell-1)}(i)) = f^{(\ell-1)}(f(i)).$$
Consequently, $\Card(f^*(f(i))) = \Card(f^*(i)) - 1$.
\end{lemma}
\begdim
The proof is by induction on $\ell \geq 1$.
Clearly $f^{(1)}(i) = f (f^{(0)}(i)) = f(i) = f^{(0)}(f(i))$.
Let $\ell \geq 2$. Then, by using induction hypothesis,
\begin{eqnarray*}
f^{(\ell)}(i) &=& f (f^{(\ell-1)}(i))
              = f ( f (f^{(\ell-2)}(i)))
              = f (f^{(\ell-2)}(f(i)))
              = f^{(\ell-1)}(f(i))
\end{eqnarray*}
and the proof is complete.
\enddim

The following result is proved in \cite[Lemma 32.5] {CLR}
({\it Prefix-function iteration lemma}).

\begin{lemma} \label{iterationlemma}
Let $w$ be a word of length $n$ with failure function $f$.
For any $i$, $1 \leq i \leq n$, we have
$f^*(i)  = \{k ~|~ k < i \mbox{ and } w[1..k] \mbox{ is a suffix of } w[1..i] \}$.
\end{lemma}

\begin{lemma} \label{prefix-property}
Let $w$ be a word of length $n$ with failure function $f$.
For any $i$, $1 \leq i \leq n$,
$f^*(i)$ lists the sequence of the lengths of all the words that are prefixes
and suffixes of $w[1..i]$ in decreasing order.
That is,
\begin{itemize}
\item[(1)]
a word $x$ is a prefix and a suffix of $w[1..i]$ if and only if
there is $k$, $0 \leq k \leq \Card(f^*(i)) - 1$, such that
$x = w[1..f^{(k)}(i)]$.
\item[(2)]
If $k < k'$,
then $|w[1..f^{(k')}(i)]| <  |w[1..f^{(k)}(i)]|$
and $w[1..f^{(k')}(i)]$ is a prefix and a suffix
of $w[1..f^{(k)}(i)]$.
\end{itemize}
\end{lemma}
\begdim
(1) Let $x$ be a prefix and a suffix of $w[1..i]$.
We prove that there is $k$, $0 \leq k \leq \Card(f^*(i)) - 1$, such that
$x = w[1..f^{(k)}(i)]$ by induction on $i$.
If $i = 1$, then $x = w[1]$ or $x = 1$.
Since $f^*(i)  = \{ 1, 0 \}$
the statement is proved for $i = 1$.
Assume $i > 1$. If $x = w[1..i] = w[1..f^{(0)}(i)]$, then we
have done. Otherwise, by the definition of the failure function,
$x$ is a prefix and a suffix of $w[1..f(i)]$.
Of course $f(i) < i$, hence, by induction hypothesis,
there is $k$, $0 \leq k \leq \Card(f^*(f(i))) - 1$, such that
$x = w[1..f^{(k)}(f(i))]$. Thus, by Lemma \ref{comm},
there is $k + 1$, $0 \leq k + 1 \leq \Card(f^*(i)) - 1$, such that
$x = w[1..f^{(k+1)}(i)]$.
Conversely, by Lemma \ref{iterationlemma}, if
there is $k$, $0 \leq k \leq \Card(f^*(i)) - 1$, such that
$x = w[1..f^{(k)}(i)]$, then $x$ is a prefix and a suffix of $w[1..i]$.

(2) By the definition of the failure function, one has
that the sequence $f^*(i)$ is strictly decreasing, that is,
if $k < k'$, with $0 \leq k , k' \leq \Card(f^*(i)) - 1$,
then $f^{(k')}(i) < f^{(k)}(i)$ and consequently
$|w[1..f^{(k')}(i)]| <  |w[1..f^{(k)}(i)]|$.
Therefore $w[1..f^{(k')}(i)]$ is a prefix and a suffix
of $w[1..f^{(k)}(i)]$ by Lemma \ref{basic4}.
\enddim

The longest proper prefix of $w[1..i]$ which is a suffix of
$w[1..i]$ is also called a {\it border} of $w[1..i]$ \cite{lothaire}.
It is known that there is an algorithm that outputs
the array $\Fi$ of $n = |w|$ integers such that $\Fi[i] = f(i)$ is the length of
the border of $w[1..i]$ in time $\mathcal{O}(n)$
(see \cite{CLR} for
a description of this procedure, called \textbf{Compute-Prefix-Function}
or \cite{lothaire}, where it is called \textbf{Border}).
Moreover, by Lemma \ref{basic4}, the array $\Fi$ determines the set $f^*(n) \setminus \{n\} =
\{ f(n), f^{(2)}(n), \ldots , f^{(m)}(n) \}$. Indeed, $\Fi[n] = f(n)$
and, if $f^k(n) = \Fi[i] = f(i) = j$, then $f^{k+1}(n) = f(f^k(n)) = f(j) = \Fi[j]$.

\begin{example} \label{bordo} {\rm
Let us consider again $w=bbabbabbb$ as in Example \ref{t}.
We have that
$\Fi=[0,1,0,1,2,3,4,5,2]$. }
\end{example}

\subsection{Description of Find-bre} \label{DescFindbre}

In this section
we present \textbf{Find-bre}, which
applies to the output $(x,y)$ of \textbf{Find-prefix}$(w)$
when $w$ is not an inverse Lyndon word.
In this case, $x = p \overline{p}$,
where $(p,\overline{p}) \in \Pref_{bre}(w)$.
As already said, the task of \textbf{Find-bre}$(x,y)$
is to find the shortest $r$ such that $x = raurb$,
where $r,u \in \Sigma^*$, $a,b \in \Sigma$, $a < b$.
Hence, by Lemma \ref{shortest}, we have
$p = rau$ and $\bar{p} = rb$.
Therefore, \textbf{Find-bre}$(x,y)$ computes the prefix $p$
and its bounded right extension $\overline{p}$
and outputs the quadruple $(p,\overline{p},y,|r|)$.
\textbf{Find-bre} uses the array $\Fi$ computed by \textbf{Border}$(raur)$.

\begin{algorithm}[htb!]
\SetKwInOut{PRE}{Input}
\SetKwInOut{POST}{Output}
\PRE{A pair of strings $(x,y)$, where $w = xy$ is not an inverse Lyndon word,
$x = p \bar{p} = raurb$, with $(p, \bar{p}) \in \Pref_{bre}(w)$, $n=|raur| = |x| - 1$.
The array $\Fi$
computed by \textbf{Border}$(raur)$.
}
\POST{A quadruple $(x_1, x_2, y, x_3)$, where $(x_1, x_2, y, x_3) = (rau, rb, y, |r|)$.}

 $i \gets n$\;
 $\LAST \gets n + 1$\;

\While{$i>0$}
 {
 \If{$w[f(i)+1] < b$}
{
$\LAST=f(i)$\;
}
 $i \gets f(i)$\;
 }
  \Return ($w[1..n-\LAST], w[n- \LAST +1, n]b, y, \LAST$) \;
\caption{Find-bre}
\label{alg1}
\end{algorithm}

\medskip
\begin{example} \label{overlap}
\rm{Let $\Sigma = \{a, b \}$ with $a < b$, let $w=bbabbabbb$.
In Example \ref{t} we noticed that \textbf{Find-prefix}$(bbabbabbb)$ outputs $(bbabbabbb,1)$.
We can check that \textbf{Find-bre}$(bbabbabbb,1)$ returns $(bbabba,bbb,1,2)$.}
\end{example}

\subsection{Correctness of Find-bre} \label{CorrectFindbre}

In this section we prove that \textbf{Find-bre} does what it is claimed to do.
To begin, notice that each time around the while-loop of lines (3) to (6),
$i$ decreases since $f(i) < i$. Thus when $i$ becomes zero, the loop
condition $i > 0$ will be false and the loop will terminate.
Then, consider the following loop-invariant statement.

\begin{quote}
{\it S(t)}:
If we reach the loop test ``$i > 0$''
with
the variable $i$ having the value
$h$ and
the variable $\LAST$ having the value $k$, after $t$ iterations of the while-loop,
then
\begin{itemize}
\item
$h = f^{t}(n)$.
\item
$k \geq h$.
Precisely, if $w[h+1] < w[n+1]$, then $k = h$ else $k > h$.
\end{itemize}
\end{quote}

\begin{proposition} \label{InizMaint2}
For any $t \geq 0$, $S(t)$ is true.
\end{proposition}
\begdim
{\bf (Basis)} Let us prove that $S(t)$ is true for $t = 0$. We reach the test after $0$
iterations of the while-loop only when we enter the loop from the outside.
Prior to the loop, lines (1) and (2) set $i$ to $n = f^{0}(n)$ and $\LAST$ to $n+1$.
Hence, we reach the test after $0$ iterations of the while-loop with
$h = f^{0}(n)$ and $k > h$.
Thus, since $w[h+1] = w[n+1]$, clearly $S(0)$ is true.

{\bf (Induction)} We suppose that $S(t)$ is true and prove that $S(t+1)$ is true.
Therefore, after $t$ iterations of the while-loop,
we reach the loop test ``$i > 0$'', with the variable $i$ having the value
$h = f^{t}(n)$. We may assume $h > 0$
(otherwise we break out of the while-loop after $t$ iterations or earlier
and $S(t + 1)$ is clearly true, since it is a conditional expression with a false antecedent).
Let us consider what happens when we run the $(t+1)$th iteration of the while-loop and
we execute the body of the while-loop
with $i$ having the value $h$ and $\LAST$ having the value $k$.

The variable $i$ assumes value $f(h) = f^{t +1}(n)$ on line (6).
Moreover, if $w[f(h)+1] = w[f^{t +1}(n) + 1] < w[n+1]$,
then $\LAST$ assumes value $f^{t +1}(n)$ (lines (4)-(5)).
Otherwise $\LAST$ remains unchanged, hence $k \geq h = f^{t}(n) > f^{t +1}(n)$
(induction hypothesis). In both cases, $S(t+1)$ is true.
\enddim

\begin{proposition} \label{termination2}
Let $w \in \Sigma^+$ be a word which is not an inverse Lyndon word.
Algorithm \textbf{Find-bre}, applied to the output $(x,y)$ of \textbf{Find-prefix}$(w)$,
outputs the quadruple $(p, \overline{p}, y, |r|)$, where $w = xy = p \overline{p} y$,
$(p,\overline{p}) \in \Pref_{bre}(w)$, $p = rau$, $\overline{p} = rb$.
\end{proposition}
\begdim
Let $w \in \Sigma^+$ be a word which is not an inverse Lyndon word.
By Proposition \ref{termination}, if
$(x,y)$ is the output of \textbf{Find-prefix}$(w)$, then
$x = p \overline{p}$, where $(p,\overline{p}) \in \Pref_{bre}(w)$,
and $w = p \overline{p}y$. Let $z$ be the prefix of $x$
such that $|z| = |x| - 1$. By Lemma \ref{shortest}, it suffices to
prove that \textbf{Find-bre} outputs
$(rau,rb, y, |r|)$, where $r$ is the shortest prefix and suffix
of $z$ such that $zb = x = p \overline{p} = raurb$,
with $r,u \in \Sigma^*$, $a, b \in \Sigma$, $a < b$.

Of course after $m = \Card(f^*(n)) - 1$ iterations, the while-loop of lines (3) to (6)
terminates. Recall also that the sequence $f^*(n)$ is strictly decreasing
(Lemma \ref{prefix-property}).
Let $r$ be the shortest prefix and suffix
of $z$ such that $z = raur$ and $w[|r|+1] = a < b = w[n+1]$.
By Lemma \ref{prefix-property}, there is $q \geq 0$ such that
$|r| = f^q(n)$.
Moreover, at line (6), the integer $t$ such that $i = f^t(n)$
is incremented by 1 each time around the loop.
Therefore, after $q$ iterations of the while-loop we have
$i = |r| = f^{q}(n)$.
By Proposition \ref{InizMaint2}, $S(q)$ is true.
Thus, $i = |r| = f^{q}(n) = \LAST$ since $w[|r|+1] = w[i+1] = a < b = w[n+1]$.
This value of $\LAST$ remains unchanged until we break out of the while-loop
since, otherwise, for $s > q$, we would have $w[f^{s}(n)+1] = c < w[n+1]$
and there would exist a shorter word $r' = w[1..f^{s}(n)]$ such that $z = r'u'r'$,
where $u'$ starts with $c$, a contradiction.
Finally, \textbf{Find-bre} outputs $(w[1..n-\LAST], w[n- \LAST +1.. n]b, y, \LAST)$.
Now $w[n- \LAST +1.. n]$ is the suffix of $z$ of length $\LAST = |r|$,
hence $w[n- \LAST +1.. n] = r$. Of course $w[1..n-\LAST] =rau$.
\enddim


\section{Computing $\ICFL$ in linear time} \label{AlgoICFL}

In this section we give a linear time algorithm, called \textbf{Compute-ICFL}
to compute $\ICFL(w)$. For the sake of simplicity we present a recursive version of
\textbf{Compute-ICFL}. In this case the correctness of the algorithm easily follows
from the definition of $\ICFL$. The output of \textbf{Compute-ICFL}
is represented as a list denoted by $list$.

Let us describe the high-level structure of algorithm
\textbf{Compute-ICFL}$(w)$. The algorithm firstly calls
\textbf{Find-prefix}$(w)$ (line (1)), that, in view of Proposition \ref{termination},
allows us to state whether $w$ is an inverse Lyndon word
or not.

If $w$ is an inverse Lyndon word, then \textbf{Find-prefix}$(w)$
returns $(x,y) = (w\$, 1)$ and \textbf{Compute-ICFL} stops
and returns $(w)$ (lines (2)-(3)), according to Definition \ref{def:ICFL}.
If $w$ is not an inverse Lyndon word, then \textbf{Find-prefix}$(w)$
returns the pair $(x,y)$ such that
$w = xy = p \overline{p} y$, $(p , \overline{p}) \in \Pref_{bre}(w)$, and
\textbf{Compute-ICFL} calls \textbf{Find-bre}$(x,y)$ (line (4)).
In turn, \textbf{Find-bre}$(x,y)$
returns a quadruple $(x_1, x_2, y, \LAST)$, where
$x_1 = p$, $x_2 = \overline{p}$ and $\LAST=|r|$.
Next, \textbf{Compute-ICFL} recursively calls itself
on $x_2y$ (line (5)) and returns $list = \ICFL(x_2y) = \ICFL(\bar{p} y)$.
Let $z = m'_1$ be the first element of $list$ (line (6)).
According to Definition \ref{def:ICFL}, we have to test
whether $x_2 = \overline{p} = rb \leq_{p} m'_1 = z$,
that is if $|z| > |r| = \LAST$. This is done on line (7).
If $|z| > |r| = \LAST$, then
we add $x_1 = p$ at the first position of $list$ (line (8)),
otherwise we replace $z = m'_1$ in $list$
with $x_1z = p m'_1$ (line (10)). In both cases,
\textbf{Compute-ICFL} returns $list = \ICFL(w)$.

It is worth of noting that there is no preprocessing of $w$ for
computing the failure function used by \textbf{Find-bre}$(x,y)$.
Each call to \textbf{Find-bre}$(x,y)$
calls \textbf{Border}$(x')$, where $(x,y)$ is the output of
\textbf{Find-prefix}$(w)$ and $x'$ is the prefix of $x$ of length
$|x| - 1$.

\begin{algorithm}[htb!]
\SetKwInOut{PRE}{Input}
\SetKwInOut{POST}{Output}
\PRE{A word $w$.}
\POST{$\ICFL(w)$}

$(x,y) \gets$ Find-prefix$(w)$\;
\If{ $x= w\$$}
{
\Return  $(w)$\;
}

$(x_1, x_2, y, \LAST)  \gets$ Find-bre$(x,y)$\;
list  $\gets$ Compute-ICFL($x_2y$)\;
$z \gets$ the first element of list\;
\If{$|z|> \LAST$}
{
 add $x_1$  in front of the list\;
 }
\Else{
 replace $z$ in list with  $x_1 z$\;
}
\Return list\;

\caption{Compute-ICFL}
\label{Algo1}
\end{algorithm}

\begin{example} \label{esecuzione}
\rm{Let $\Sigma = \{a, b, c \}$ with $a < b < c$,
let $w=cbabacaacbabacbac$,
already considered in Example \ref{Ex1}.
Suppose that we call \textbf{Compute-ICFL} on $w$ and on the empty list $list$.
The three tables below illustrate the sequence of calls made to
\textbf{Compute-ICFL}, \textbf{Find-prefix} and \textbf{Find-bre}
if we read the first column downward.
For instance, since \textbf{Find-prefix}$(cbac)$ returns $(cbac\$, 1)$,
\textbf{Compute-ICFL}$(cbac)$ returns $(cbac)$ without invoking itself
again and the recursion stops.
\textbf{Compute-ICFL}$(cbabacbac)$ calls \textbf{Find-prefix} on
$cbabacbac$, \textbf{Find-bre} on $(cbabacbac, 1)$ and then
\textbf{Compute-ICFL}$(cbac)$ which returns $(cbac)$.
Since $|z| = |cbac| = 4 > \LAST = 3$,
\textbf{Compute-ICFL}$(cbabacbac)$ returns $(cbaba,cbac)$.
Finally, \textbf{Compute-ICFL}$(w)$ calls \textbf{Find-prefix} on
$w$, \textbf{Find-bre} on $(cbabacaacbabacb,ac)$ and then
\textbf{Compute-ICFL}$(cbabacbac)$ which returns $(cbaba,cbac)$.
Since $|z| = |cbaba|= 5 \leq \LAST = 6$, \textbf{Compute-ICFL}$(w)$
replaces $cbaba$ with the concatenation of $cbabacaa$ and $cbaba$ and returns
$list=(cbabacaacbaba,cbac)=\ICFL(w)$.}

\medskip

\centerline
{\begin{tabular}{|c|c|}\hline
CALL & RETURN  \\ \hline
\textbf{Compute-ICFL}$(cbabacaacbabacbac)$ & $(cbabacaacbaba,cbac)$ \\ \hline
\textbf{Compute-ICFL}$(cbabacbac)$ & $(cbaba,cbac)$ \\ \hline
\textbf{Compute-ICFL}$(cbac)$ & $(cbac)$ \\ \hline
\end{tabular}
}

\bigskip

\centerline
{\begin{tabular}{|c|c|}\hline
CALL & RETURN  \\ \hline
\textbf{Find-prefix}$(cbabacaacbabacbac)$ & $(cbabacaacbabacb,ac)$ \\ \hline
\textbf{Find-prefix}$(cbabacbac)$ & $(cbabacbac,1)$ \\ \hline
\textbf{Find-prefix}$(cbac)$ & $(cbac\$, 1)$ \\ \hline
\end{tabular}
}

\bigskip

\centerline
{\begin{tabular}{|c|c|}\hline
CALL & RETURN  \\ \hline
\textbf{Find-bre}$(cbabacaacbabacb, ac)$ & $(cbabacaa, cbabacb, ac, 6)$ \\ \hline
\textbf{Find-bre}$(cbabacbac, 1)$ & $(cbaba, cbac, 1, 3)$ \\ \hline
\end{tabular}
}
\end{example}

\subsection{Performance of \textbf{Compute-ICFL}}

Let us compute the running time  $T(n)$ of \textbf{Compute-ICFL} when $w$ has length $n$.
We can check that the running time of \textbf{Find-prefix}$(w)$ is
$\mathcal{O}(|x|)$, when the output of the procedure  is $(x, y)$.
Then, \textbf{Compute-ICFL} calls the procedure \textbf{Find-bre}$(x,y)$,
where $x = p \overline{p}$ and $(p,\overline{p}) \in \Pref_{bre}(w)$.
By Lemma \ref{lemma0}, we know that $|\overline{p}| \leq |p|$, and so
the running time of \textbf{Find-bre} is $\mathcal{O}(|p| + |\overline{p}|) = \mathcal{O}(|p|)$.
Let $\ICFL(w) = (m_1, \ldots , m_k)$ and let $n_j$ be the length of $m_j$,
$1 \leq j \leq k$. The recurrence for $T(n)$ is defined as
$T(n) = T(n - n_1) + \mathcal{O}(n_1)$, where $T(n_k)$ is $\mathcal{O}(n_k)$,
because there is no recursive call in this case.
It is easy to see that the solution to this recurrence is
$T(n) = \sum_{j = 1}^{k} \mathcal{O}(n_j) = \mathcal{O}(n)$,
since $w = m_1 \cdots  m_k$.


\section{Groupings} \label{groupings}

Let $(\Sigma, <)$ be a totally ordered alphabet. As we know,
for any word $w \in \Sigma^+$, there are three sequences of words associated with $w$:
the Lyndon factorization of $w$ with respect to the order
$\prec$, denoted $\CFL(w)$, the Lyndon factorization of $w$
with respect to the inverse lexicographic order $\prec_{in}$, denoted $\CFL_{in}(w)$ and the inverse Lyndon
factorization $\ICFL(w)$ of $w$. In this section, we compare
$\CFL_{in}(w)$ and $\ICFL(w)$.

We begin by proving some relations between inverse Lyndon words
and anti-Lyndon words (Section \ref{inv}).
Then, we point out some relations between $\ICFL(w)$ and $\CFL_{in}(w)$.
Precisely, starting with $\CFL_{in}(w)$, we define a family of inverse Lyndon
factorizations of $w$, called {\it groupings} of $\CFL_{in}(w)$
(Section \ref{relations}).
We prove that $\ICFL(w)$ is a grouping of $\CFL_{in}(w)$ in
Section \ref{icfl-grouping}. Then, we prove that
the set of the inverse Lyndon words is equal to the set
of the strict sesquipowers of anti-Lyndon words (Section \ref{SSA}).
Finally, we prove that
Theorem \ref{teo-sorting} may be generalized to groupings
when we refer to the sorting with respect to the inverse
lexicographic order (Section \ref{sortingsuffixes}).

\subsection{Inverse Lyndon words and anti-Lyndon words} \label{inv}

Let $(\Sigma, <)$ be a totally ordered alphabet, let $<_{in}$ be the inverse of $<$
and let $\prec_{in}$ be the inverse lexicographic order on $(\Sigma, <)$.
The following proposition justifies the adopted terminology.

\begin{proposition} \label{P3}
Let $(\Sigma, <)$ be a totally ordered alphabet.
For all $x, y \in \Sigma^*$ such that $x \Join y$,
$$y \prec_{in} x \Leftrightarrow x \prec y .$$
Moreover, in this case $x \ll y$.
\end{proposition}
\begdim
Let $x, y \in \Sigma^*$ such that $x \Join y$.
Assume $y \prec_{in} x$. Thus, by Definition \ref{ILO}, there are
$a,b \in \Sigma$, with $b <_{in} a$, and $r,s,t \in \Sigma^*$ such that
$y = rbs$, $x =rat$. Hence $a < b$ and $x \prec y$, by Definition \ref{lex-order}.
Conversely, if $x \prec y$ we have $y \prec_{in} x$
by a similar argument. The second part of the statement follows by
Definition \ref{lex-order}.
\enddim

Lemma \ref{propinvlexord} is a dual version of item (2) in Lemma \ref{proplexord}.

\begin{lemma} \label{propinvlexord}
If $y \prec_{in} x$ and $x \Join y$, then $yu \prec_{in} xv$
for all words $u,v$.
\end{lemma}
\begdim
If $y \prec_{in} x$ and $x \Join y$, then $x \ll y$, by Proposition \ref{P3}.
Then, by item (2) in Lemma \ref{proplexord}, $xv \ll yu$ for all words $u,v$.
Since $xv \Join yu$, the conclusion follows by Proposition \ref{P3}.
\enddim

The following proposition characterizes the set
$L_{in} = L_{(\Sigma^*, <_{in})}$ of the anti-Lyndon words on $\Sigma^*$.

\begin{proposition} \label{Lyforinv}
A word $w \in \Sigma^+$ is in $L_{in}$ if and only if $w$ is primitive and
$w \succ vu$, for each $u, v \in \Sigma^+$ such that $w = uv$.
\end{proposition}
\begdim
By Definition \ref{Lyndon-word}, if $w \in L_{in}$, then $w$ is nonempty and primitive.
Moreover, if $w = uv$, with $u,v \not = 1$, then $w \prec_{in} vu$.
Since $uv \Join vu$, by Proposition \ref{P3} one has $vu \prec w$, i.e.,
$w \succ vu$.
A similar argument shows that if $w$ is a primitive nonempty word
and $w \succ vu$, for each $u, v \in \Sigma^+$ such that $w = uv$,
then $w \in L_{in}$.
\enddim

We state below a slightly modified dual version of Proposition \ref{P2}. It shows that
anti-Lyndon words are inverse Lyndon words.

\begin{proposition} \label{P4}
A word $w \in \Sigma^+$ is in $L_{in}$ if and only if $w$ is unbordered
and $w \succ v$, for each proper nonempty suffix $v$.
\end{proposition}
\begdim
Let $w \in L_{in} = L_{(\Sigma^*, <_{in})}$. Therefore, $w$ is nonempty and unbordered (Definition
\ref{Lyndon-word}, Proposition \ref{P1}).
Moreover, if $w = uv$, with $u,v \not = 1$, then $w \prec_{in} v$, by Proposition \ref{P2}.
In addition, $w \Join v$ since $w$ is unbordered and $|v| < |w|$.
Consequently, $w \succ v$ (Proposition \ref{P3}).

Conversely, let $w = uv$ a nonempty unbordered word such that $w \succ v$, for each
proper nonempty suffix $v$. Thus, $v \Join w$, hence $w \prec_{in} v$ (Proposition \ref{P3}).
By Proposition \ref{P2}, the word $w$ is in $L_{in}$.
\enddim

Of course there are inverse Lyndon words which are not anti-Lyndon words.
For instance consider $\Sigma = \{a,b\}$, with
$a < b$. The word $bab$ is an inverse Lyndon words but it is not unbordered, thus
it is not an anti-Lyndon word.

The following result give more precise relations between words in $L_{in}$
and their proper nonempty suffixes.

\begin{proposition} \label{P9}
If $v$ is a proper nonempty suffix of $w \in L_{in}$, then $v \ll w$.
\end{proposition}
\begdim
Let $v$ be a proper nonempty suffix of $w \in L_{in}$.
By Proposition \ref{P4}, we have $v \prec w$.
Moreover, since $w$ is unbordered, we have $v \Join w$, hence $v \ll w$.
\enddim

Some compositional properties of the inverse Lyndon words are proved below.

\begin{proposition} \label{P5}
For any $w \in L_{in}$ and $h \geq 1$,
the word $w^h$ is an inverse Lyndon word on $(\Sigma^*, <)$.
\end{proposition}
\begdim
Let $w \in L_{in}$ and let $vw^i$ be a proper nonempty suffix of $w^h$, $h \geq 1$.
If $v = 1$, then $0< i < h$, hence $w^i$ is a nonempty proper prefix of $w^h$.
By Definition \ref{lex-order}, $w^h \succ w^i$.
Otherwise, $v$ is a proper nonempty suffix of $w$. Thus, by Proposition \ref{P9},
$v \ll w$ and by item (2) in Lemma \ref{proplexord}, $vw^i \ll w^h$, i.e., $w^h \succ vw^i$.
\enddim

\begin{proposition} \label{P6}
Let $\ell_1, \ell_2 \in L_{in}$. If
$\ell_2$ is a proper prefix of $\ell_1$,
then $\ell_1\ell_2$ is an inverse Lyndon word.
\end{proposition}
\begdim
Let $\ell_1,\ell_2$ be as in the statement.
Set $\ell_1 \ell_2 = \ell_2 x \ell_2 = w$.
If $s$ is a nonempty proper suffix of $w$, then one of the following cases holds
\begin{itemize}
\item[(1)]
$s$ is a nonempty proper suffix of $\ell_2$
\item[(2)]
$s = \ell_2$
\item[(3)]
$s = s'\ell_2$, where $s'$ is a nonempty proper suffix of $\ell_1$.
\end{itemize}
Assume that case (1) holds. Thus $s \ll \ell_2$, by Proposition \ref{P9}.
Consequently, $s \ll \ell_2(x \ell_2) = w$, by
item (2) in Lemma \ref{proplexord}.
Of course, $w = \ell_2 x \ell_2 \succ \ell_2$ (case (2)), therefore assume that case (3) holds,
i.e., $s = s'\ell_2$, where $s'$ is a nonempty proper suffix of $\ell_1$.
Arguing as before, $s' \ll \ell_1$ (Proposition \ref{P9}), hence
$s' \ell_2 \ll \ell_1 \ell_2 = w$ (item (2) in Lemma \ref{proplexord}).
\enddim

\begin{example}
{\rm Let $\ell_1, \ldots , \ell_h$ be words in $L_{in}$ which form
a non-increasing chain $\ell_1 \geq_p \ldots  \geq_p \ell_h$ for the prefix order,
i.e.,
$\ell_i$ is a prefix of $\ell_{i-1}$, $1 < i \leq h$.
The word $\ell_1 \cdots  \ell_h$ is not necessarily an inverse Lyndon word.
As an example, let $\Sigma = \{a, b \}$ with $a < b$. The sequence $baa, ba, b$
is such that $baa \geq_p ba \geq_p b$. The word
$baabab$ is not an inverse Lyndon word since
$baabab \ll bab$.}
\end{example}

\subsection{A family of inverse Lyndon factorizations of words} \label{relations}

Groupings of $\CFL_{in}(w)$ are special inverse Lyndon factorizations.
They are constructed in a very natural way. We first give some needed definitions
and results.

\begin{definition} \label{MaxCh}
Let $w \in \Sigma^+$, let $\CFL_{in}(w) = (\ell_1, \ldots , \ell_h)$
and let $1 \leq r < s \leq h$.
We say that $\ell_{r}, \ell_{r+1} \ldots , \ell_{s}$
is a non-increasing {\rm maximal chain for the prefix order}
in $\CFL_{in}(w)$, abbreviated $\PMCI$, if
$\ell_{r} \geq_p \ell_{r+1} \ldots  \geq_p \ell_{s}$.
Moreover, if $r > 1$, then $\ell_{r - 1} \not \geq_p \ell_{r}$,
if $s < h$, then $\ell_{s} \not \geq_p \ell_{s +1}$.
Two $\PMCI$ $\mathcal{C}_1 = \ell_{r}, \ell_{r+1} \ldots , \ell_{s}$,
$\mathcal{C}_2 = \ell_{r'}, \ell_{r'+1} \ldots , \ell_{s'}$ are
{\rm consecutive} if $r' = s+1$ (or $r = s' +1$).
\end{definition}

\begin{lemma} \label{intermedioinverseLyndon1}
Let $x,y$ be nonempty words such that $x \succeq_{in} y$.
Then either $x \geq_p y$ or $x \ll y$.
\end{lemma}
\begdim
Let $x,y$ be nonempty words such that $x \succeq_{in} y$.
Therefore $x$ is not a proper prefix of $y$. If $y$ is not a prefix of $x$,
then $x \Join y$ and $y \prec_{in} x$. Hence, by Proposition \ref{P3}, we have
$x \ll y$.
\enddim

The following is a direct consequence of Lemma \ref{intermedioinverseLyndon1}.

\begin{proposition} \label{canCFL}
Let $w \in \Sigma^+$, let $\CFL_{in}(w) = (\ell_1, \ldots , \ell_h)$.
Then
$$(\ell_1, \ldots , \ell_h) =
(\mathcal{C}_1, \ldots ,\mathcal{C}_t),$$
where any
$\mathcal{C}_j$, $1 \leq j \leq t$,
is a $\PMCI$
in $\CFL_{in}(w)$. Moreover, $\mathcal{C}_j$, $\mathcal{C}_{j+1}$
are consecutive and $\ell \ll \ell'$, where $\ell$ is the last word in $\mathcal{C}_j$
and $\ell'$ is the first word in $\mathcal{C}_{j+1}$,
for $1 \leq j \leq t-1$.
\end{proposition}
\begdim
Let $w \in \Sigma^+$ and let $\CFL_{in}(w) = (\ell_1, \ldots , \ell_h)$, i.e.,
\begin{eqnarray} \label{ILF1}
w & = & \ell_1  \cdots \ell_h, \mbox{ with } \ell_j \in L_{in} \mbox{ and } \ell_1 \succeq_{in}  \ldots \succeq_{in} \ell_h
\end{eqnarray}
By Lemma \ref{intermedioinverseLyndon1},
each symbol $\succeq_{in}$ in Eq. (\ref{ILF1}) may be replaced
either by $\geq_p$ or by $\ll$. Therefore, the conclusion follows.
\enddim

The definition of a grouping of $\CFL_{in}(w)$ is given below in two steps. We first
define the grouping of a $\PMCI$. Then a grouping of $\CFL_{in}(w)$ is obtained
by changing each $\PMCI$ with one of its groupings.

\begin{definition} \label{grouping1}
Let $\ell_1, \ldots , \ell_h$ be words in $L_{in}$ such that
$\ell_i$ is a prefix of $\ell_{i-1}$, $1 < i \leq h$.
We say that $(m_1, \ldots , m_k)$ is a {\rm grouping} of $(\ell_1, \ldots , \ell_h)$
if the following conditions are satisfied.
\begin{itemize}
\item[(1)]
$m_j$ is an inverse Lyndon word which is a product of consecutive $\ell_q$, $1 \leq j \leq k$.
\item[(2)]
$\ell_1 \cdots  \ell_{h} = m_1 \cdots m_k$,
\item[(3)]
$m_1 \ll \ldots \ll m_k$.
\end{itemize}
\end{definition}

We now extend Definition \ref{grouping1} to $\CFL_{in}(w)$.

\begin{definition} \label{grouping2}
Let $w \in \Sigma^+$ and let $\CFL_{in}(w) = (\ell_1, \ldots , \ell_h)$.
We say that $(m_1, \ldots , m_k)$ is a {\rm grouping} of $\CFL_{in}(w)$ if
it can be obtained by replacing
any $\PMCI$ $\mathcal{C}$ in $\CFL_{in}(w)$
by a grouping of $\mathcal{C}$.
\end{definition}

Proposition \ref{main2} shows that groupings of $\CFL_{in}(w)$
are inverse Lyndon factorizations of $w$.
However, as Example \ref{esempiogrouping1} shows, there are
inverse Lyndon factorizations which are not groupings.

\begin{example} \label{esempiogrouping1}
{\rm Let $\Sigma = \{a,b,c,d \}$ with $a < b < c < d$, and
let $w = dabadabdabdadac$.
Thus, $d <_{in} c <_{in} b <_{in} a$ and
$\CFL_{in}(w) = (daba, dab, dab, dadac)$.
The two sequences $(daba, dabdab, dadac)$,
$(dabadab, dabda, dac)$ are both inverse Lyndon factorizations of $w$ and
moreover $\ICFL(w) = (daba, dabdab, dadac)$
(Examples \ref{nonunique}, \ref{Ex2}).
However, $\ICFL(w)$ is a grouping of $\CFL_{in}(w)$, whereas
$(dabadab, dabda, dac)$ is not a grouping of $\CFL_{in}(w)$.
In Example \ref{nonunique} we also
considered the following two inverse Lyndon factorizations
of $z = dabdadacddbdc$
$$(dab)(dadacd)(db)(dc) = (dabda)(dac)(ddbdc)$$
Both of them are not groupings since
$\CFL_{in}(z) = (dab,dadac,ddbdc)$.
Notice that $\CFL_{in}(z) = \ICFL(z)$ (see Corollary \ref{Pr1}).}
\end{example}

\begin{proposition} \label{main2}
Let $w \in \Sigma^+$. If $(m_1, \ldots , m_k)$ is a grouping of $\CFL_{in}(w)$,
then $(m_1, \ldots , m_k)$ is an
inverse Lyndon factorization of $w$.
\end{proposition}
\begdim
Let $w \in \Sigma^+$ and let $\CFL_{in}(w) = (\ell_1, \ldots , \ell_h)$.
Therefore, by Proposition \ref{canCFL}, we have
$$(\ell_1, \ldots , \ell_h) =
(\mathcal{C}_1, \ldots ,\mathcal{C}_t),$$
where any
$\mathcal{C}_j$, $1 \leq j \leq t$,
is a $\PMCI$
in $\CFL_{in}(w)$. Moreover, $\mathcal{C}_j$, $\mathcal{C}_{j+1}$
are consecutive and $\ell \ll \ell'$, where $\ell$ is the last word in $\mathcal{C}_j$
and $\ell'$ is the first word in $\mathcal{C}_{j+1}$,
for $1 \leq j \leq t-1$.

Now let $(m_1, \ldots , m_k)$ be a
grouping of $\CFL_{in}(w)$.
Any $m_j$ is an inverse Lyndon word since it is an element of a grouping of a
$\PMCI$ in $\CFL_{in}(w)$.
Then, let $\mathcal{S}_j$ be the product of the words in
$\mathcal{C}_j$, $1 \leq j \leq t$.
It is clear that
$$w = \ell_1 \cdots  \ell_h = \mathcal{S}_1 \cdots \mathcal{S}_t
= m_1, \ldots , m_k.$$
Finally, set $\mathcal{C}_j = (\ell_i, \ldots , \ell_g)$
and $\mathcal{C}_{j+1} = (\ell_{g+1}, \ldots , \ell_f)$,
for $1 \leq j \leq t-1$.
Thus $\ell_g \ll \ell_{g+1}$.
It suffices to show that if $(m_r, \ldots , m_s)$ is the grouping
of $\mathcal{C}_j$ that replaces $\mathcal{C}_j$
and $(m_{s+1}, \ldots , m_v)$ is the grouping
of $\mathcal{C}_{j+1}$ that replaces $\mathcal{C}_{j+1}$,
then $m_s \ll m_{s+1}$.
But this is clear since $\ell_g$ is a suffix of $m_s$, hence it is also a prefix of $m_s$ ($\mathcal{C}_j$
is a $\PMCI$) and $\ell_{g+1}$ is a prefix of $m_{s+1}$. Therefore, by item (2) in Lemma \ref{proplexord},
$\ell_g \ll \ell_{g+1}$ implies $m_s \ll m_{s+1}$.
\enddim

\subsection{$\ICFL(w)$ is a grouping of $\CFL_{in}(w)$} \label{icfl-grouping}

Let us outline how we prove below that $\ICFL(w)$ is a grouping of
$\CFL_{in}(w)$.
Let $\ICFL(w) = (m_1, \ldots , m_k)$, let $\CFL_{in}(w) = (\ell_1, \ldots , \ell_h)$
and let $\ell_1, \ldots , \ell_q$ be a $\PMCI$ in $\CFL_{in}(w)$, $1 \leq q \leq k$.
The proof will be divided into four steps.
\begin{itemize}
\item[(1)]
We prove that $m_1$ cannot be a proper prefix of $\ell_1$ (Proposition \ref{Pr2}).
\item[(2)]
We prove that if $m_1$ is a prefix of
$\ell_1 \cdots  \ell_{q}$, then $m_1 = \ell_1 \cdots  \ell_{q'}$,
for some $q'$, $1 \leq q' \leq q$
(Proposition \ref{Pr3}).
\item[(3)]
We prove that $\ell_1 \cdots  \ell_{q}$ cannot be a proper prefix
of $m_1$ (Proposition \ref{Pr4}).
\item[(4)]
We complete the proof by induction on $|w|$ (Proposition \ref{Pr5}).
\end{itemize}
We say that a sequence of nonempty words $(m_1, \ldots , m_k)$ is a {\it factorization}
of $w$ if $w = m_1 \cdots m_k$.
It is worth of noting that steps (1) and (2) are proved under the more general hypothesis
that $(m_1, \ldots , m_k)$ is a factorization
of $w$ such that $m_1 \ll m_2$ and, for step (2), where $m_1$ is an inverse Lyndon word.
Proposition \ref{Pr0} and Corollary \ref{Pr1}
deal with two extremal cases where
there is only one grouping of $\CFL_{in}(w)$, namely $\ICFL(w)$.

\begin{proposition} \label{Pr0}
Let $(\Sigma, <)$ be a totally ordered alphabet. Let $w \in \Sigma^+$ and let
$\CFL_{in}(w) = (\ell_1, \ldots , \ell_h)$.
If $w$ is an inverse Lyndon word, then either $w$ is unbordered or
$\ell_1, \ldots , \ell_{h}$ is a $\PMCI$ in $\CFL_{in}(w)$.
In both cases $\ICFL(w) = (w)$ is the unique grouping of $\CFL_{in}(w)$.
\end{proposition}
\begdim
Let $w \in \Sigma^+$ be an inverse Lyndon word and let
$\CFL_{in}(w) = (\ell_1, \ldots , \ell_h)$.
We know that $\ICFL(w) = (w)$ (Definition \ref{def:ICFL}).
By Proposition \ref{P4}, if
$w$ is unbordered, then $w$ is an anti-Lyndon word.
Thus, by item (iii) in Lemma \ref{duval-prop},
$\CFL_{in}(w) = (w)$
and of course this is the unique grouping of $\CFL_{in}(w)$.

Otherwise, $w$ is bordered and, again by Proposition \ref{P4}, $h > 1$.
By contradiction assume that $\ell_1, \ldots , \ell_{h}$ is not a $\PMCI$ in $\CFL_{in}(w)$.
By Lemma \ref{intermedioinverseLyndon1}, there would be a smallest $q$,
$1 \leq q \leq h-1$ such that
$$\ell_1 \geq_p \ldots  \geq_p \ell_{q} \ll \ell_{q+1}$$
Hence, since $\ell_{q}$ is a prefix of $w$, by item (2) in Lemma \ref{proplexord}, we would have
$w \ll  \ell_{q+1} \cdots \ell_{h}$, which is a contradiction
since $w$ is an inverse Lyndon word and $\ell_{q+1} \cdots \ell_{h}$
is a proper nonempty suffix of $w$. Thus,
$\ell_1, \ldots , \ell_{h}$ is a $\PMCI$ in $\CFL_{in}(w)$ and
$\ICFL(w) = (w)$ is a grouping of $\CFL_{in}(w)$.
By contradiction, assume that
$(m_1, \dots , m_k)$, $k \geq 2$, is another grouping of $\CFL_{in}(w)$.
Therefore, $m_1 \ll m_2$ and by item (2) in Lemma \ref{proplexord}, we would have
$w \ll  m_2 \cdots m_k$, which is a contradiction
since $w$ is an inverse Lyndon word and $m_2 \cdots m_k$
is a proper nonempty suffix of $w$.
\enddim

\begin{proposition} \label{Pr2}
Let $(\Sigma, <)$ be a totally ordered alphabet. Let $w \in \Sigma^+$ and let
$\CFL_{in}(w) = (\ell_1, \ldots , \ell_h)$.
For any factorization $(m_1, \ldots , m_k)$
of $w$, with $k > 1$, if $m_1 \ll m_2$, then $m_1$ cannot be a proper prefix
of $\ell_1$.
\end{proposition}
\begdim
Let $(m_1, \ldots , m_k)$ be a factorization of $w$ such that
$m_1 \ll  m_2$.
By contradiction assume that $m_1$ is a proper prefix of $\ell_1$.
Therefore, there are two nonempty words $x,y$ such that $m_1 = x$, $\ell_1 = xy$,
and $v \in \Sigma^*$ such that $m_{2} \cdots m_k = yv$.
On one hand, by Proposition \ref{P9} we have $y \ll \ell_1$.
On the other hand, $x = m_1 \ll m_{2}$, thus $x = m_1 \ll m_2 \cdots m_k= yv$,
by item (2) in Lemma \ref{proplexord}.
Consequently, $x = ras$, $yv = rbt$, with $a < b$. Hence either $y$ is a prefix of
$r$, which is a contradiction since $\ell_1 = xy = rasy$ is unbordered (Proposition \ref{P4}),
or $rb$ is a prefix of $y$, which is once again a contradiction
since we would have $x \ll y$ and consequently, by item (2) in Lemma \ref{proplexord},
$\ell_1 \ll y$.
\enddim

\begin{proposition} \label{Pr3}
Let $(\Sigma, <)$ be a totally ordered alphabet. Let $w \in \Sigma^+$ and let
$\CFL_{in}(w) = (\ell_1, \ldots , \ell_h)$.
Let $(m_1, \ldots , m_k)$ be a factorization
of $w$ such that $k > 1$, $m_1$ is an inverse Lyndon word and $m_1 \ll m_2$.
Let $\ell_1, \ldots , \ell_q$ be a $\PMCI$ in $\CFL_{in}(w)$, $1 \leq q \leq h$.
If $m_1$ is a prefix of
$\ell_1 \cdots  \ell_{q}$, then $m_1 = \ell_1 \cdots  \ell_{q'}$,
for some $q'$, $1 \leq q' \leq q$.
\end{proposition}
\begdim
Let $w \in \Sigma^+$ and let
$\CFL_{in}(w) = (\ell_1, \ldots , \ell_h)$.
Let $(m_1, \ldots , m_k)$ be a factorization
of $w$ such that $k > 1$, $m_1$ is an inverse Lyndon word and $m_1 \ll m_2$.
Let
$\ell_1, \ldots , \ell_{q}$ be a $\PMCI$ in $\CFL_{in}(w)$, $1 \leq q \leq h$.

By contradiction, assume that there are two nonempty words $x,y$
such that $m_1 = \ell_1 \cdots \ell_{j-1}x$, $xy = \ell_{j}$, $1 \leq j \leq q$
(where it is understood that $m_1 = x$ when $j = 1$).
By Proposition \ref{Pr2}, we have $j > 1$. Moreover there is
$v \in \Sigma^*$ such that $m_{2} \cdots m_k = yv$.

Since $m_1 \ll m_2$, we have
$m_1 = \ell_1 \cdots \ell_{j-1}x \ll m_2 \cdots m_k = yv$.
Hence, there are words $r,s, t \in \Sigma^*$ and letters $a,b \in \Sigma$,
with $a < b$ such that
$m_1 = ras$, $m_2  \cdots m_k = yv = rbt$.
If $|r| < |y|$, then $rb$ is a prefix of $y$, therefore it is a factor
of $\ell_{j}$. In turn, $\ell_{j}$ is a prefix of $\ell_{j-1}$,
thus there is $\gamma \in \Sigma^*$ such that $rb \gamma$ is a proper nonempty suffix of $m_1$.
Then, $m_1 = ras \ll rb \gamma$, a contradiction, since $m_1$ is an
inverse Lyndon word.
Hence $|r| \geq |y|$, i.e., $y$ is a prefix of $r$ and consequently it
is a prefix of $m_1$.
The word $\ell_{j} = xy$ is also a prefix of $m_1$,
thus $y$ and $\ell_{j} = xy$ are comparable for the prefix order.
Therefore $y$ is both a nonempty proper prefix and a suffix of $\ell_{j}$,
i.e., $\ell_{j} = xy$ is bordered,
a contradiction since $\ell_{j} \in L_{in}$
(see Proposition \ref{P4}).
\enddim

\begin{proposition} \label{Pr4}
Let $(\Sigma, <)$ be a totally ordered alphabet. Let $w \in \Sigma^+$, let
$\CFL_{in}(w) = (\ell_1, \ldots , \ell_h)$ and let $\ICFL(w) = (m_1, \ldots , m_k)$.
Let
$\ell_1, \ldots , \ell_{q}$ be a
$\PMCI$ in $\CFL_{in}(w)$, $1 \leq q \leq h$.
Then $m_1 = \ell_1 \cdots  \ell_{q'}$,
for some $q'$, $1 \leq q' \leq q$.
\end{proposition}
\begdim
Let $w \in \Sigma^+$, let
$\CFL_{in}(w) = (\ell_1, \ldots , \ell_h)$, and let
$\ICFL(w) = (m_1, \ldots , m_k)$.
We prove the statement by induction on $|w|$.
If $|w| = 1$, then $w$ is an inverse Lyndon word and we have done, by
Proposition \ref{Pr0}.
Hence assume $|w| > 1$.
If $w$ is an inverse Lyndon word,
then the proof is ended, once again by Proposition \ref{Pr0}.
Therefore, assume that $w$ is not an inverse Lyndon word.

Let
$\ell_1, \ldots , \ell_{q}$ be a
$\PMCI$ in $\CFL_{in}(w)$, $1 \leq q \leq h$.
Since $m_1$ and $\ell_1 \cdots  \ell_{q}$ are both prefixes of $w$,
$m_1$ and $\ell_1 \cdots  \ell_{q}$ are comparable for the prefix order.
By contradiction,
assume that $m_1$ violates the statement.
Therefore, by Propositions \ref{Pr2} and \ref{Pr3},
$m_1$ is not a prefix of $\ell_1 \cdots  \ell_{q}$.
Hence, $\ell_1 \cdots  \ell_{q}$ is a proper prefix of $m_1$, i.e.,
there are two words $x,y$ and $j$, with $q < j \leq h$,
such that $m_1 = \ell_1 \cdots \ell_{j-1}x$, $xy = \ell_{j}$.
Moreover, $\ell_{q} \ll \ell_{q+1}$
and, if $j-1 = q$, then $x \not = 1$.
We must have $j - 1 = q$ (thus $x \not = 1$ also) and $y \not = 1$.
Indeed, otherwise $j - 1 > q$ or $x = \ell_{j}$, $j \geq q+1$.
In both cases, $\ell_{q + 1} \cdots \ell_{j-1}x$ would be a proper nonempty suffix
of $m_1$ and $\ell_{q}$ is a prefix of $\ell_1$, hence $\ell_{q}$
is a prefix of $m_1$.
By item (2) in Lemma \ref{proplexord} applied to $\ell_{q} \ll \ell_{q+1}$,
we would have $m_1 \ll \ell_{q + 1} \cdots \ell_{j-1}x$,
a contradiction since $m_1$ is an inverse Lyndon word.
In conclusion, there are words $x, y, y'$, with $x \not = 1$ and $y \not = 1$, such that
\begin{eqnarray} \label{EQC1}
&& m_1 = \ell_1 \cdots \ell_{q}x, \quad \ell_{q + 1} = xy, \quad m_2 \cdots m_k = yy',
\quad \ell_1 \geq_p \ldots  \geq_p \ell_{q} \ll \ell_{q+1}
\end{eqnarray}
Let $(p, \bar{p}) \in \Pref_{bre}(w)$. Let
$v \in \Sigma^+$ be such that $w = pv$ and let
$\ICFL(v) = (m'_1, \ldots , m'_{k'})$.
By Definition \ref{def:ICFL}, one of the following two cases holds
\begin{itemize}
\item[(1)]
$m_1 = p$
\item[(2)]
$m_1 = pm'_1$.
\end{itemize}
In both cases, $p$ is a prefix of $\ell_1 \cdots \ell_{q}x$, therefore
$|p| \leq |\ell_1 \cdots \ell_{q}x|$. We claim that we also have
$|p| > |\ell_1 \cdots \ell_{q}|$. By Eq. (\ref{EQC1}), this is clearly true
in case (1), hence assume $m_1 = pm'_1$.
Since $p \ll \bar{p}$, by item (2) in Lemma \ref{proplexord}, we have
$p \ll v$. Thus, Propositions \ref{Pr2} and \ref{Pr3} apply to the factorization
$(p, v)$ of $w$. Therefore, if $|p| \leq |\ell_1 \cdots \ell_{q}|$, then
$p = \ell_1 \cdots \ell_{j}$, with $j \leq q$. Hence,
$v = \ell_{j + 1} \cdots  \ell_h$ and, by Theorem \ref{Lyndon-factorization},
$\CFL_{in}(v) = (\ell_{j + 1}, \ldots , \ell_h)$.
On the other hand, by Eq. (\ref{EQC1}), we would have
$m'_1 = \ell_{j + 1} \cdots \ell_{q}x$, $\ell_{q + 1} = xy$, $\ell_{q} \ll \ell_{q+1}$,
and $x \not = 1$,
in contradiction with induction hypothesis applied to $v$.
In conclusion, in both cases (1) and (2), we have
$|\ell_1 \cdots \ell_{q}| < |p| \leq |\ell_1 \cdots \ell_{q}x|$, thus
there are words $x_1,x_2$ such that
\begin{eqnarray} \label{EQC2}
&& p = \ell_1 \cdots \ell_{q}x_1, \;
x = x_1x_2, \; x_1 \not = 1, \;
x_2 = \begin{cases} 1 & \text{if $m_1 = p$}, \\
m'_1 & \text{if $m_1 = pm'_1$} \end{cases}
\end{eqnarray}
Then, by Definitions \ref{brepref} and \ref{def:ICFL},
there are words $r,s, t \in \Sigma^*$ and letters $a,b \in \Sigma$,
with $a < b$ such that
\begin{eqnarray} \label{EQC3}
&& p = ras, \quad \bar{p} = rb, \quad v = x_2m_2 \cdots m_k = rbt
\end{eqnarray}
Since $\ell_{q} \ll \ell_{q+1}$, there are words $z,f,g \in \Sigma^*$ and letters $c,d \in \Sigma$,
with $c < d$ such that
\begin{eqnarray} \label{EQC4}
&& \ell_{q} = zcf, \quad \ell_{q+1} = xy = zdg
\end{eqnarray}
Observe that $x$ and $zd$ are both prefixes of $\ell_{q+1}$, hence they are
comparable for the prefix order.
If $zd$ would be a prefix of $x$, then for a word $\gamma$,
$zd \gamma$ would be a proper nonempty suffix of $m_1$ (see Eq. (\ref{EQC1}))
and $\ell_{q}$ would be a prefix of $\ell_1$, thus of $m_1$
such that $\ell_{q} \ll zd \gamma$ (see Eq. (\ref{EQC4})).
By item (2) in Lemma \ref{proplexord} applied to $\ell_{q} \ll zd \gamma$,
we would have $m_1 \ll zd \gamma$,
a contradiction since $m_1$ is an inverse Lyndon word.
Therefore, $x$ is a proper prefix of $zd$, i.e., there is $z' \in \Sigma^*$
such that
\begin{eqnarray} \label{EQC5}
&& z = xz' = x_1x_2z'.
\end{eqnarray}
Eqs. (\ref{EQC4}) and (\ref{EQC5}) yield $xy = zdg = xz'dg$, therefore
$z'$ is a prefix of $y$ and thus
$x_2z'$ is a prefix of $x_2m_2 \cdots m_k$ (see Eq. (\ref{EQC1})).
On the other hand $rb$ is also a prefix of $x_2m_2 \cdots m_k$ (see Eq. (\ref{EQC3})).
Hence, $rb$ and $x_2z'$ are comparable for the prefix order and
one of the following two cases is satisfied
\begin{itemize}
\item[(i)]
$rb$ is a prefix of $x_2z'$,
\item[(ii)]
$x_2z'$ is a prefix of $r$.
\end{itemize}
Assume that case (i) holds. In this case,
since $x_2z'$ is a suffix of $z$ (Eq. (\ref{EQC5})) and $z$
is a prefix of $\ell_{q}$ (Eq. (\ref{EQC4})), the word $rb$ would be a factor
of $\ell_{q}$.  Thus, by Eq. (\ref{EQC1}),
there would be $\gamma \in \Sigma^*$ such that $rb \gamma$ would be a proper nonempty
suffix of $m_1$.
Since $ras = p$ (Eq. (\ref{EQC3})), the
word $ras$ would be a prefix of $m_1$ and we would have $m_1 \ll rb \gamma$, a contradiction
since $m_1$ is an inverse Lyndon word.
Case (ii) also leads to a contradiction.
Indeed, set $p' = \ell_1 \cdots \ell_{q}$.
Then, by Eqs. (\ref{EQC2}) and (\ref{EQC5}),
$px_2z' = p'x_1x_2z' = p'z$.
If $x_2z'$ is a prefix of $r$, then $px_2z' = p'z$ is a prefix of
$pr$, thus $px_2z' = p'z$ is a proper prefix of $p \bar{p}$ (Eq. (\ref{EQC3})),
hence $px_2z' = p'z$
is an inverse Lyndon word, as do
all nonempty prefixes of $p'z$, by Definition \ref{brepref}.
Moreover, by Eq. (\ref{EQC4}), $zc$ is a prefix of $\ell_{q}$, thus of $p'$.
Therefore, we have $p' = zcf'$, for a word $f'$, and $p'zd$ is not an inverse
Lyndon word since $p'zd \ll zd$. On the contrary,
$zd$ is an inverse Lyndon word, because
it is a prefix of $\ell_{q+1}$ (Eq. \ref{EQC4}).
Finally, $p' = zcf' \ll zd$.
By Definition \ref{brepref}, the pair $(p', zd) \in \Pref_{bre}(w)$.
Since $x_1 \not = 1$, we have $p' \not = p = p' x_1$, in contradiction with
Proposition \ref{unica}.
\enddim

\begin{proposition} \label{Pr5}
Let $(\Sigma, <)$ be a totally ordered alphabet. For any $w \in \Sigma^+$,
$\ICFL(w)$ is a grouping of $\CFL_{in}(w)$.
\end{proposition}
\begdim
Let $w \in \Sigma^+$, let
$\CFL_{in}(w) = (\ell_1, \ldots , \ell_h)$, and let
$\ICFL(w) = (m_1, \ldots , m_k)$.
The proof is by induction on $|w|$.
If $|w| = 1$, then $w$ is an inverse Lyndon word and we have done, by
Proposition \ref{Pr0}.
Hence assume $|w| > 1$.

If $w$ is an inverse Lyndon word, once again by Proposition \ref{Pr0},
$\ICFL(w) = (w) = (\ell_1 \cdots  \ell_h)$ is a
grouping of $\CFL_{in}(w)$.
Therefore, assume that $w$ is not an inverse Lyndon word.
By Propositions \ref{Pr2}-\ref{Pr4}, there is $j$,
$1 \leq j \leq h$ such that $m_1 = \ell_1 \cdots  \ell_j$,
where $\ell_1 \geq_p \ldots  \geq_p \ell_{j}$.

Let $(p, \bar{p}) \in \Pref(w)_{bre}$. Let $\ICFL(v) = (m'_1, \ldots , m'_{k'})$,
where $v \in \Sigma^+$ is such that $w = pv$.
By induction hypothesis, $(m'_1, \ldots , m'_{k'})$ is a grouping of
$\CFL_{in}(v)$.
By Definition \ref{def:ICFL}, one of the following two cases holds
\begin{itemize}
\item[(1)]
$m_1 = p$, $(m_2, \ldots , m_k) = (m'_1, \ldots , m'_{k'})$, i.e., $k = k' + 1$, $m_j = m'_{j -1}$, $2 \leq j \leq k$.
\item[(2)]
$m_1 = pm'_1$, $(m_2, \ldots , m_k) = (m'_2, \ldots , m'_{k'})$, i.e., $k = k'$, $m_j = m'_j$, $2 \leq j \leq k$.
\end{itemize}

(Case (1)). In this case,
since $p = m_1 =\ell_1 \cdots  \ell_j$, we have $v = \ell_{j+1} \cdots  \ell_h$,
where any $\ell_g$ is in $L_{in}$ and
$\ell_{j+1} \succeq_{in} \ldots \succeq_{in} \ell_h$.
By Theorem \ref{Lyndon-factorization},
$\CFL_{in}(v) = (\ell_{j+1}, \ldots , \ell_h)$.
Consequently, if $(\ell_{j+1}, \ldots , \ell_h) =
(\mathcal{C}_1, \ldots ,\mathcal{C}_t)$, where any
$\mathcal{C}_j$, $1 \leq j \leq t$, is a $\PMCI$ in $\CFL_{in}(v)$,
then $(\ell_1, \ldots , \ell_h) =
(\mathcal{C}, \mathcal{C}_1, \ldots ,\mathcal{C}_t)$
where $\mathcal{C} =  \ell_1, \ldots , \ell_j$,
any $\mathcal{C}_j$, $2 \leq j \leq t$, is a $\PMCI$
in $\CFL_{in}(w)$ and either $\mathcal{C}, \mathcal{C}_1$ represents
two $\PMCI$ in $\CFL_{in}(w)$ or it represents a single
$\PMCI$ in $\CFL_{in}(w)$.
In both cases, since $m_1 = \ell_1 \cdots  \ell_j$
and given that $(m_2, \ldots , m_k) = (m'_1, \ldots , m'_{k'})$ is a grouping of
$\CFL_{in}(v)$, we conclude that
$\ICFL(w)$ is a grouping of $\CFL_{in}(w)$.

(Case (2)). Set
$\CFL_{in}(v) = (\ell'_1, \ldots , \ell'_{h'})$, and
let $j'$ be such that
$m'_1 = \ell'_{1} \cdots  \ell'_{j'}$,
$\ell'_{1} \geq_p \ldots  \geq_p \ell'_{j'}$.
Therefore we have
\begin{eqnarray} \label{EQG1}
(pm'_1) (m_2 \cdots  m_k) & =& (\ell_1 \cdots  \ell_j) (\ell_{j+1} \cdots  \ell_h)
=  (p \ell'_{1} \cdots  \ell'_{j'})( \ell'_{j'+ 1} \cdots  \ell'_{h'})
\end{eqnarray}
Since $(pm'_1) = (\ell_1 \cdots  \ell_j) = (p \ell'_{1} \cdots  \ell'_{j'})$,
Eq. (\ref{EQG1}) implies
\begin{eqnarray} \label{EQG2}
(m_2 \cdots  m_k) & = & (\ell_{j+1} \cdots  \ell_h)
= (\ell'_{j'+ 1} \cdots  \ell'_{h'})
\end{eqnarray}
All the words $\ell_g, \ell'_{g'}$ are in $L_{in}$ and
$\ell_{j+1} \succeq_{in} \ldots \succeq_{in} \ell_h$,
$\ell'_{j'+1} \succeq_{in} \ldots \succeq_{in} \ell'_{h'}$.
Hence, by Theorem \ref{Lyndon-factorization}, applied to
$v' = m_2 \cdots m_k = m'_2 \cdots  m'_{k'}$,
Eq. (\ref{EQG2}) implies $(\ell_{j+1}, \ldots  , \ell_h)
= (\ell'_{j'+ 1},  \ldots  , \ell'_{h'})$ and
$\CFL_{in}(v') = (\ell_{j+1}, \ldots , \ell_h)$.
Now, $(m'_1, \ldots , m'_{k'})$ is a grouping of
$\CFL_{in}(v)$, where $v = m'_1v'$ and
$m'_1 = \ell'_{1} \cdots  \ell'_{j'}$.
Thus, $(m_2, \ldots , m_h) = (m'_2, \ldots , m'_{k'})$
is a grouping of $(\ell_{j+1}, \ldots  , \ell_h)
= (\ell'_{j'+ 1},  \ldots  , \ell'_{h'}) = \CFL_{in}(v')$.
The rest of the proof runs as in Case (1).
Namely, if $(\ell_{j+1}, \ldots , \ell_h) =
(\mathcal{C}_1, \ldots ,\mathcal{C}_t)$, where any
$\mathcal{C}_j$, $1 \leq j \leq t$, is a $\PMCI$ in $\CFL_{in}(v')$, then
$(\ell_1, \ldots , \ell_h) =
(\mathcal{C}, \mathcal{C}_1, \ldots ,\mathcal{C}_t)$
where $\mathcal{C} =  \ell_1, \ldots , \ell_j$,
any $\mathcal{C}_j$, $2 \leq j \leq t$, is a $\PMCI$
in $\CFL_{in}(w)$ and either $\mathcal{C}, \mathcal{C}_1$ represents
two $\PMCI$ in $\CFL_{in}(w)$ or it represents a single
$\PMCI$ in $\CFL_{in}(w)$.
In both cases, since $m_1 = \ell_1 \cdots  \ell_j$
and given that $(m_2, \ldots , m_k) = (m'_2, \ldots , m'_{k'})$ is a grouping of
$\CFL_{in}(v')$, we conclude that
$\ICFL(w)$ is a grouping of $\CFL_{in}(w)$.
\enddim

\begin{corollary} \label{Pr1}
Let $(\Sigma, <)$ be a totally ordered alphabet. Let $w \in \Sigma^+$ and let
$\CFL_{in}(w) = (\ell_1, \ldots , \ell_h)$, with $h > 1$.
If $\ell_1 \ll  \ldots \ll \ell_h$, then $\ICFL(w) = \CFL_{in}(w)$ and this
is the unique grouping of $\CFL_{in}(w)$.
\end{corollary}
\begdim
Let $w \in \Sigma^+$ and let
$\CFL_{in}(w) = (\ell_1, \ldots , \ell_h)$, with $h > 1$.
If $\ell_1 \ll  \ldots \ll \ell_h$, then $\CFL_{in}(w)$
is the unique grouping of $\CFL_{in}(w)$ (Definition \ref{grouping2}).
Thus, by Proposition \ref{Pr5}, $\ICFL(w) = \CFL_{in}(w)$.
\enddim

The following example shows that $\ICFL(w)$ is in general different from
$\CFL_{in}(w)$.

\begin{example}
{\rm Let $\Sigma = \{a,b \}$ with $a < b$,
let $w = bab \in \Sigma^+$.
Therefore, $b <_{in} a$ and
$\CFL_{in}(w) = (ba, b)$. Since $bab$ is an inverse Lyndon word, we have
$\ICFL(w) = (bab)$.}
\end{example}

The following example shows that there are words $w$ such that $\CFL_{in}(w)$
has more than one grouping, thus there are groupings of $\CFL_{in}(w)$ different from
$\ICFL(w)$.

\begin{example} \label{ExCom}
{\rm Let $\Sigma = \{a,b,c,d \}$ with $a < b < c < d$,
let $w = dabadabdabdabdadac \in \Sigma^+$.
Therefore, $d <_{in} c <_{in} b <_{in} a$.
Moreover, $\CFL_{in}(w) = (daba, dab, dab, dab, dadac)$.
The two sequences $(dabadab, (dab)^2, dadac)$,
$(daba, (dab)^3, dadac)$ are both groupings of $\CFL_{in}(w)$.
Let us compute $\ICFL(w)$.
The shortest prefix of $w$
which is not an inverse Lyndon word is $dabadabd$
and $(daba, dabd) \in \Pref_{bre}(w)$ by Lemma \ref{shortest}.
Set $w = dabay$, we have to compute $\ICFL(y)$.
The shortest prefix of $y = (dab)^3dadac$
which is not an inverse Lyndon word is $(dab)^3dad$
and $((dab)^3, dad) \in \Pref_{bre}(y)$ by Lemma \ref{shortest}.
On the other hand, since $dadac$ is an inverse Lyndon word,
we have $\ICFL(dadac) = (dadac)$ and thus
$\ICFL(y) = ((dab)^3, dadac)$ (see Definition \ref{def:ICFL}).
Finally, by Definition \ref{def:ICFL},
$\ICFL(w) = (daba, (dab)^3, dadac)$.
}
\end{example}

We end the section with Proposition \ref{S3}. It  shows that nonempty words which are both proper prefix and
suffix of an inverse Lyndon word have a special form.

\begin{proposition} \label{S3}
Let $(\Sigma, <)$ be a totally ordered alphabet. Let $w \in \Sigma^+$ be an inverse Lyndon word and let
$\CFL_{in}(w) = (\ell_1, \ldots , \ell_h)$. For any proper nonempty suffix $u$ of
$w$ such that $u$ is also a prefix of $w$, there is $j_u$ such that $u = \ell_{j_u} \cdots \ell_h$,
$2 \leq j_u \leq h$.
\end{proposition}
\begdim
Let $(\Sigma, <)$ be a totally ordered alphabet. Let $w \in \Sigma^+$ be a
bordered inverse Lyndon word and let
$\CFL_{in}(w) = (\ell_1, \ldots , \ell_h)$.
Let $u \in \Sigma^+$ be such that $u$ is a proper nonempty suffix of $w$ and
$u$ is also a prefix of $w$.
By contradiction, assume that $u = s \ell_{j_u} \cdots \ell_h$,  $2 \leq j_u \leq h$,
where $s$ is a nonempty proper suffix of $\ell_{j_u -1}$.
Since $w$ is a bordered inverse Lyndon word, by Proposition \ref{Pr0},
$\ell_1, \ldots , \ell_{h}$ is a $\PMCI$ in $\CFL_{in}(w)$.
Hence $\ell_{j_u -1}$ is a prefix of $w$. It follows that $\ell_{j_u -1}$ and $s$ are comparable
for the prefix order and consequently $\ell_{j_u -1}$ is bordered, in contradiction with
$\ell_{j_u -1} \in L_{in}$.
\enddim

\subsection{Strict sesquipowers of anti-Lyndon words} \label{SSA}

For a word $w \in \Sigma^*$ and a letter $a \in \Sigma$, we denote by $|w|_a$ the number of
occurrences of $a$ in $w$. Then, the set
$$\al(w)  = \{ a \in \Sigma ~|~ |w|_a > 0 \}$$
is the set of all letters occurring at least once in $w$.

Let $\PAL = \{w \in \Sigma^+ ~|~ w\Sigma^* \cap L_{in} \not = \emptyset \}$ be
the set of the nonempty prefixes of anti-Lyndon words, let
$a$ be the minimum letter in $\Sigma$ and let $P' = \PAL \cup \{a^m ~|~ m \geq 2 \}$.
Let $\SP$ be the set of strict sesquipowers of the words in $L_{in}$, that is
$$\SP =\{(uv)^nu ~|~ u \in \Sigma^*, v \in  \Sigma^+, n \geq 1 \mbox{ and } uv \in L_{in} \}.$$
The following result has been proved in \cite{duval}.

\begin{proposition} \label{SP}
$$\SP = P' = \PAL \cup \{a^m ~|~ m \geq 2 \}.$$
\end{proposition}

The aim of this section is to prove that $\SP$ is equal to the class of the
inverse Lyndon words.

\begin{proposition} \label{S1}
For any $a \in \Sigma$ and $m \geq 1$, the word $a^m$ is an inverse Lyndon word.
\end{proposition}
\begdim
By definition, any letter $a \in \Sigma$ is in $L_{in}$. Therefore,
by Proposition \ref{P5}, the word $a^m$ is an inverse Lyndon word, for $m \geq 1$.
\enddim

\begin{proposition} \label{S2}
Let $(\Sigma, <)$ be a totally ordered alphabet. Let $w \in \Sigma^+$ be an inverse Lyndon
word. If $w$ starts with the minimum letter in $\Sigma$, say $a$, then $w = a^m$, with $m \geq 1$.
\end{proposition}
\begdim
Let $(\Sigma, <)$ be a totally ordered alphabet and let $a$ be
the minimum letter in $\Sigma$. Let $w,w'$ be words such that $w = a w'$ is
an inverse Lyndon word. By contradiction, let $w_1,w_2$ be words such that
$w = aw_1bw_2$, with $b \not = a$. Hence $b > a$ and $w \ll bw_2$, in contradiction
with the hypothesis.
\enddim

As another step in our proof, we show that any inverse Lyndon word can be extended
as a word which is still an inverse Lyndon word.

\begin{proposition} \label{S4}
Let $(\Sigma, <)$ be a totally ordered alphabet and let $a$ be the minimum letter in $\Sigma$.
For any inverse Lyndon word $w$ and for any nonnegative integer
$n$, the word $wa^n$ is an inverse Lyndon word.
\end{proposition}
\begdim
Let $(\Sigma, <)$ be a totally ordered alphabet and let $a$ be the minimum letter in $\Sigma$.
Let $w$ be an inverse Lyndon word.

If $w$ starts with $a$, then $w = a^m$, with $m \geq 1$, by
Proposition \ref{S2}. Thus, by Proposition \ref{S1}, the word $wa^n = a^{m+n}$
is still an inverse Lyndon word.

Now, let $b \in \Sigma$, $w' \in \Sigma^*$ be such that $w = bw'$, with $b > a$.
We first prove that $wa$ is an inverse Lyndon word, that is, $wa \succ s$, for any
proper nonempty suffix of $wa$.
Any proper nonempty suffix $s$ of $wa$ has the form $s = s'a$, where $s'$ is a proper
suffix of $w$. If $s' = 1$, then $a = s \ll wa = bw'a$
and we have done. If $s' \not = 1$, then $w \succ s'$ because $w$ is an inverse Lyndon word.
In addition, if $s' \ll w$, then $s = s'a \ll wa$, by item (2) in Lemma \ref{proplexord}.
Otherwise, $s'$ is both a prefix and a proper nonempty suffix of $w$, hence $w = s'y$
for a nonempty word $y$.
By contradiction, assume $s' \prec wa = s'ya \prec s'a$. By item (3) in Lemma \ref{proplexord},
$ya \prec a$ which implies $ya \ll a$, a contradiction since $a$ is the
minimum letter in $\Sigma$.
The proof is completed by using induction. Indeed, if $wa^n$ is an inverse Lyndon word
which starts with $b$,
the above argument shows that $wa^{n+1}$ is still an inverse Lyndon word
which starts with $b$.
\enddim

We prove below that any inverse Lyndon word can be extended
as a word which is still an inverse Lyndon word and which is additionally unbordered.
Proposition \ref{S5} shows that this can be easily obtained if we do not limit the size of
the alphabet whereas Proposition \ref{S6} states the result for a fixed alphabet.
Let $(\Sigma_1, <_1), (\Sigma_2, <_2)$ be totally ordered alphabets such that
$\Sigma_1 \subseteq \Sigma_2$. We say that $<_2$ is an {\it extension} of $<_1$
if the following holds
$$ \forall a, b \in \Sigma_1 \quad a <_{1} b \Leftrightarrow a <_2 b .$$
If $<_2$ is an extension of $<_1$, then inverse Lyndon words in $\Sigma_1^+$
are also inverse Lyndon words in $\Sigma_2^+$.

\begin{proposition} \label{S5}
Let $(\Sigma_1, <_1)$ be a totally ordered alphabet.
For any inverse Lyndon word $w \in \Sigma_1^+$, there is a totally ordered alphabet
$(\Sigma_2, <_2)$, where $\Sigma_1 \subseteq \Sigma_2$ and
$<_2$ is an extension of $<_1$, such that $w \in \SP$.
\end{proposition}
\begdim
Let $(\Sigma_1, <_1)$ be a totally ordered alphabet and let
$w \in \Sigma_1^+$ be an inverse Lyndon word.
Let $\alpha$ be a letter such that $\alpha \not \in \Sigma_1$
and set $\Sigma_2 = \Sigma_1 \cup \{\alpha\}$.
Let $(\Sigma_2, <_2)$ be the totally ordered alphabet, where
$<_2$ is the extension of $<_1$ defined as follows
\begin{eqnarray*}
\forall a \in \Sigma_1 && \alpha <_2 a , \\
\forall a, b \in \Sigma_1 && a <_{1} b \Leftrightarrow a <_2 b.
\end{eqnarray*}
By Proposition \ref{S4}, $w \alpha$ is an inverse Lundon word.
Moreover, $w \alpha$ is unbordered, thus $w \alpha \in L_{in}$ (Proposition \ref{P4})
and consequently $w \in \PAL \subseteq \SP$.
\enddim

\begin{proposition} \label{S6}
Let $(\Sigma, <)$ be a totally ordered alphabet and let $w \in \Sigma^+$.
If $w$ is an inverse Lyndon word, then $w \in \SP$.
\end{proposition}
\begdim
Let $(\Sigma, <)$ be a totally ordered alphabet and let $w \in \Sigma^+$
be an inverse Lyndon word. Let $a$ be the minimum letter in $\Sigma$.
If $w$ starts with $a$, then $w = a^m$, with $m \geq 1$ (Proposition \ref{S2}),
hence $w \in \SP$. Otherwise, let $b \in \Sigma$, $w' \in \Sigma^*$ be such
that $w = bw'$, with $b > a$.
If $w = bw'$ is unbordered, then $w$ is an anti-Lyndon word (Proposition \ref{P4})
and consequently $w \in \SP$. Otherwise, $b \in \al(w')$.
Let $x$ be the word that precedes the last occurrence of $b$ in $w'$, that is, let
$x, y \in \Sigma^*$ be such that $w' = xby$ and $b \not \in \al(y)$.
Let $n > |x| + 1$ and let $z = wa^n = bxbya^n$.
By Proposition \ref{S4}, the word $z$ is an inverse Lyndon word.
We claim that $z$ is unbordered, thus $z = w a^n \in L_{in}$ (Proposition \ref{P4})
and consequently $w \in \PAL \subseteq \SP$.
On the contrary, let $u$ be a proper prefix of $z = wa^n = bxbya^n$ which is also a suffix of $z$.
On one hand, $|u|_b < |z|_b$, because $u$ is a proper suffix of $z$.
On the other hand, $u$ starts with $b$ since it is a prefix of $z$.
Thus $|u| \geq |bya^n| > |x| + 2 = |bxb|$, because $b \not \in \al(ya^n)$.
Finally, since $u$ is a prefix of $z$ such that 
$|u| > |bxb|$, it follows that
$|u|_b = |z|_b$.
This contradiction ends the proof.
\enddim

\begin{proposition} \label{S7}
A word $w$ is an inverse Lyndon word if and only if $w \in \SP$.
\end{proposition}
\begdim
By Proposition \ref{S6}, if $w$ is an inverse Lyndon word, then $w \in \SP$.
Conversely, let $w \in \SP$. If $w = a^m$, $m \geq 1$, where $a$ is
(the minimum letter) in $\Sigma$, then $w$ is an inverse Lyndon word by
Proposition \ref{S1}. Otherwise, $w \in \PAL$ is a prefix
of an anti-Lyndon word, hence $w$ is an inverse Lyndon
word by Lemma \ref{lem:inverse-Lyndon-word-prefix}.
\enddim

\subsection{Sorting suffixes in $\ICFL(w)$} \label{sortingsuffixes}

In this section we use the same notation and terminology as in Section \ref{restivo-sorting}.
We prove that the same compatibility property proved in \cite{restivo-sorting-2014} holds
for the sorting of the nonempty suffixes of a word $w$ with respect to $\prec_{in}$
if we replace $\CFL(w)$ with $\ICFL(w)$.

\begin{theorem} \label{teo-sorting-inverse}
Let $w$ be a word and let $(m_1, \ldots , m_k)$ be
a grouping of $\CFL_{in}(w)$.
Then, for any $r,s$, $1 \leq r \leq s \leq k$,
the sorting with respect to $\prec_{in}$
of the nonempty local suffixes of
$w$ with respect to $u = m_r \cdots m_s$
is compatible with the sorting with respect to $\prec_{in}$
of the corresponding nonempty global suffixes of $w$.
\end{theorem}
\begdim
Let $w$ and $(m_1, \ldots , m_k)$ be as in the statement.
Let $\CFL_{in}(w) = (\ell_1, \ldots , \ell_h)$.
Let $u = m_r \cdots m_s$, with $1 \leq r \leq s \leq k$.
By Definitions \ref{grouping1}, \ref{grouping2}, any $m_j$ is a concatenation
of consecutive $\ell_q$. Hence $u$ is also a concatenation
of consecutive $\ell_q$.
By Theorem \ref{teo-sorting}, for all $i,j$ with $first(u) \leq i < j \leq last(u)$, we have
\begin{eqnarray} \label{E6}
&& suf_u(i) \prec_{in} suf_u(j) \Longleftrightarrow suf(i) \prec_{in} suf(j).
\end{eqnarray}
\enddim

The following corollary is a direct consequence of Proposition \ref{Pr5} and Theorem \ref{teo-sorting-inverse}.

\begin{corollary} \label{teo-sorting-inverseBIS}
Let $w$ be a word and let $\ICFL(w) = (m_1, \ldots , m_k)$.
Then, for any $r,s$, $1 \leq r \leq s \leq k$,
the sorting with respect to $\prec_{in}$
of the nonempty local suffixes of
$w$ with respect to $u = m_r \cdots m_s$
is compatible with the sorting with respect to $\prec_{in}$
of the corresponding nonempty global suffixes of $w$.
\end{corollary}

On the contrary, we give below a counterexample showing that the compatibility property of
local and global nonempty suffixes
does not hold in general for inverse Lyndon factorizations with respect to $\prec_{in}$
(and with respect to $\prec$).

\begin{example}
\label{ex-noncomp}
{\rm Let $(\Sigma,<)$ be as in Example \ref{nonunique}  and
let $w = daddbadc \in \Sigma^+$. Therefore, $d <_{in} c <_{in} b <_{in} a$.
Consider the inverse Lyndon factorization $(dad,  dba,  dc)$ of $w$, with
$dad \ll dba \ll dc$ and the factor $u = daddba$.
Consider the local suffixes $a, addba$ of $u$ and the corresponding
global suffixes
$adc$ and $addbadc$. We have
that $addbadc \prec_{in}  adc$ while $a \prec_{in} addba$.
Consequently, in general, the compatibility property does not hold for inverse Lyndon factorizations
with respect to $\prec_{in}$.
It does not hold also with respect to $\prec$ and even for $\ICFL$.
Indeed, let $w = dabadabdabdabdadac \in \Sigma^+$.
We know that $\ICFL(w) = (daba, (dab)^3, dadac)$ (see Example \ref{ExCom}).
For the local suffixes $dab, dabdab$ of $(dab)^3$ we have
$dab \prec dabdab$ but for the corresponding global suffixes $dabdadac, dabdabdadac$
we have $dabdabdadac \prec dabdadac$.}
\end{example}


\end{document}